\documentclass[twocolumn,secnumarabic,superscriptaddress,amssymb, nobibnotes, aps, prl, showpacs]{revtex4-1}

\usepackage[export]{adjustbox}

\usepackage[english]{babel}
\usepackage{layout}
\usepackage[top=2.5cm, bottom=2.cm, left=2.cm, right=2.cm]{geometry}
\usepackage{float}
\restylefloat{table}
\usepackage{amsthm}
\usepackage{amsmath}
\usepackage{amssymb}
\usepackage{mathrsfs}
\usepackage{stmaryrd}
\usepackage[usenames]{xcolor}
\usepackage[backref=false]{hyperref}
\hypersetup{
colorlinks=true,
citecolor=red,
linkcolor=darkblue
}
\definecolor{darkblue}{rgb}{0,0,.8}
\definecolor{red}{rgb}{1,0,0}
\usepackage{graphicx}
\usepackage{dsfont}
\usepackage{tikz}

\newcommand{\nc}{\newcommand}
\nc{\be}{\begin{equation}}
\nc{\ee}{\end{equation}}
\nc{\bea}{\begin{eqnarray}}
\nc{\eea}{\end{eqnarray}}
\nc{\bra}[1]{\langle{#1}|}
\nc{\ket}[1]{|{#1}\rangle}
\nc{\hbdy}{h_{\rm bdy}}
\nc{\hbdyvec}{{\bf h}_{\rm bdy}}

\DeclareMathOperator{\Tr}{Tr}

\nc{\iq}{\text{I}_q}
\nc{\td}{\tilde}

\begin{document}

\title{What is the group of a quantum group? \\
From $SL_q(2)$ to MPO representations of $SL(2)$}

\author{Weronika Wiesiolek}
\affiliation{Department of Applied Mathematics and Theoretical Physics, University of Cambridge,\\ Wilberforce Road, Cambridge, CB3 0WA, United Kingdom}

\author{Romain Couvreur}
\affiliation{\it Department of Physics and Astronomy, Ghent University, Krijgslaan 281, 9000 Gent, Belgium}

\author{\\Dmitry Chernyak}
\affiliation{\it Department of Physics and Astronomy, Ghent University, Krijgslaan 281, 9000 Gent, Belgium}

\author{Laurens Lootens}
\affiliation{Department of Applied Mathematics and Theoretical Physics, University of Cambridge,\\ Wilberforce Road, Cambridge, CB3 0WA, United Kingdom}

\author{Frank Verstraete}
\affiliation{Department of Applied Mathematics and Theoretical Physics, University of Cambridge,\\ Wilberforce Road, Cambridge, CB3 0WA, United Kingdom}
\affiliation{\it Department of Physics and Astronomy, Ghent University, Krijgslaan 281, 9000 Gent, Belgium}

\begin{abstract}
Generalized symmetries realized by matrix product operators (MPOs) have so far been tied to discrete data leaving continuous symmetries outside the framework. Quantum groups fill this gap: starting from the RTT relations of $SL_q(2)$ at an odd root of unity $q^d=1$, we construct a three-parameter family of periodic-boundary MPOs of bond dimension $d^2$ that multiply according to the group law of $SL(2)$. This gives Lusztig's quantum Frobenius map a concrete operator form, and endows the XXZ chain on the ring at $\Delta=(q+q^{-1})/2$ with an intrinsically non-local $SU(2)$ symmetry. Fusion of the local tensors is generically semisimple. On special loci it is non-semisimple, and yields reducible but indecomposable tensors with Jordan blocks. Differentiating $\mathcal{D}(g)$ at the identity produces the $\mathfrak{sl}(2)$ generators as $d$-body operators: the MPO algebra supplies the exponential map for quantum group algebras.
\end{abstract}

\maketitle

Symmetries are a keystone in the description of topological phases of matter, conformal field theories and integrable systems. Recently, much attention has been paid to the study of categorical symmetries, also called topological defect lines or matrix product operator (MPO) symmetries, as these symmetries reveal important non-local properties in 2d spin systems \cite{aasen2016topological,vanhove2018mapping} and anyonic spin chains \cite{gils2013anyonic,buican2017anyonic,BGJST20}. Moreover, those MPO symmetries \cite{pirvu2010matrix,schuch2010peps,bultinck2017anyons} were found to be particularly useful to characterise topological sectors and anyons in the framework of projected entangled pair states (PEPS)\cite{cirac2021matrix}.  An important class of MPO symmetries was shown to be governed by bimodule categories \cite{etingof2016tensor}, where the consistency equations associated to having MPO symmetries can be understood as a set of coupled pentagon equations with generalised $F$-symbols \cite{LFHSV}. This picture has enabled a lattice understanding of the Fuchs-Runkel-Schweigert construction of rational conformal field theory \cite{vanhove2022topological}, and provides new insights into dualities in 1D quantum models \cite{LDOV21}. It has also provided a new perspective on the problem of magnetic monopole scattering~\cite{Rubakov:1982fp,Callan:1982ah,Polchinski:fermion_rotor,Loladze:2025jsq,vanBeest:monopole}, in which a localized wavepacket is converted into a string-like object that is completely characterized by the MPO operator \cite{ueda2025perfect}.

So far, this approach has focused on discrete MPO symmetries, either based on finite groups or fusion categories.  In this paper, we extend this construction to the case of continuous groups, by taking advantage of the quantum group deformation of $SL(2)$  at roots of unity. In the seminal papers~\cite{woronowicz1987compact,faddeev1988quantization}, matrix quantum groups were defined as sets of matrices whose entries are elements of an algebra, and do not commute under multiplication. Here, this structure is reflected in the fact that the MPOs have a bond dimension larger than 1, providing the necessary non-commutative structure. These quantum groups and their relations to the algebraic Bethe ansatz are well known to describe symmetries of the XXZ chain \cite{PS90}, and to explain degeneracies in the  spectrum \cite{VJS11} and the appearance of  Fabricius-McCoy strings in the corresponding Bethe ansatz equations \cite{de1992quantum,Fabricius_2001, Fabricius:2000em, Miao_2021}.  The existence of continuous families of irreducible representations of quantum groups at roots of unity also has deep implications for the nonequilibrium physics of integrable spin chains, in particular anomalous transport properties and relaxation to non-thermal steady states after a quench \cite{PROSEN20141177,Pereira_2014, Ilievski_2016}. These peculiar dynamical features are believed to survive mild integrability breaking \cite{Brandino_2015}, potentially making our approach (which does not require integrability) particularly relevant for such systems. Quantum groups have also been put forward as genuine global symmetries of two-dimensional field theories \cite{gabai2025quantum,gabai2025quantumII}.

The MPO constructions in this paper can be viewed simultaneously as representations of two distinct structures. The local tensors form representations of the matrix quantum group $SL_q(2)$ with $q$ a primitive root of unity such that $q^{d}=1$ and $d$ is odd, while globally the MPO chains with periodic boundary conditions form a group representation of $SL(2)$.  Our construction provides an explicit operator realization, at the level of matrix product operators, of the classical group $SL(2)$ that emerges, via Lusztig's quantum Frobenius map, inside the matrix
quantum group at a root of unity \cite{Lusztig1989,Lusztig1990jams,Lusztig1990,Lusztig1993book,ParshallWang1991,DeConciniLyubashenko1994}: the labels $g$ of our MPOs are precisely the characters of the corresponding central Hopf subalgebra. By differentiating our group representations around the identity element, we recover the $\mathfrak{sl}(2)$ generators as sums of $d$-body $U_q(\mathfrak{sl}_2)$ operators; these can be understood in terms of Lusztig's divided powers $E^{(d)}$, $F^{(d)}$ \cite{Lusztig1990}, and are similar but not equal to the loop-algebra  generators built from the same divided powers in \cite{deguchi2001sl2}.  Our MPOs can in fact be read as implementing the exponential map into the quantum group algebra --- albeit multiplied by a nontrivial projector.

We will see that these MPO representations are non-semisimple, i.e. fusion of two local MPO tensors need not yield a direct sum of local MPO tensors. This work is thus an important step in understanding MPO symmetries with non-semisimple fusion rules, based on arbitrary pre-bialgebras \cite{liu2025tradingmathematicalphysicalsimplicity}, and defines novel recoupling coefficients that, to the best of our knowledge, have not been derived previously in the quantum group literature. A complementary line of attack on the non-semisimple setting is provided by the non-semisimple TQFT programme \cite{kerler2001nonsemisimple,derenzi2022threedimensional,gainutdinov2018modularization}.

\paragraph{Quantum groups and RTT relations:} Matrix quantum groups can be defined starting from the so-called RTT relations \cite{woronowicz1987compact,faddeev1988quantization,Biedenharn}. For the spin $1/2$ case, this leads to the matrix 
\begin{equation}
T=
\begin{pmatrix}
a & b \\
c & d 
\end{pmatrix}
\label{localmpodef}
\end{equation}
with entries that are themselves non-commuting operators satisfying the particular RTT relations
\begin{gather}
ab=q^{-1}b a,\quad ac=q^{-1}ca, \quad bd=q^{-1}db,\quad cd=q^{-1}dc,\nonumber\\ 
bc=cb,\quad ad-da=-(q-q^{-1})bc
\label{rttrelation}
\end{gather}
together with the following quantum determinant constraint:
\bea
ad-q^{-1}bc=\mathds{1}.
\eea
The matrix $T$ can be interpreted as the defining tensor of a uniform matrix product operator (MPO) with periodic boundary conditions, in which the physical level is 2-dimensional and the virtual level consists of the operator space generated by $\{a, b, c, d\}$. MPOs satisfying these constraints commute with the transfer matrix of the 6-vertex model or equivalently with the XXZ Hamiltonian at $\Delta=(q+q^{-1})/2$. Indeed, it was demonstrated in \cite{rubio2026} that the RTT relations above can be derived by requiring that the MPO commutes with this Hamiltonian.

For the case where $q^d=1$ with $d>1$ an odd integer, finite-dimensional $d\times d$ matrix solutions exist to the above equations. Considering the permutation matrix $\sigma$ and the diagonal matrix $\text{I}_q$  
\begin{equation}
\sigma=
\begin{pmatrix} 
0 & 1 & 0 &\cdots\\
  & 0 & 1 &   \\
& & \ddots & \\
1 &  & & 
\end{pmatrix}, \quad
\iq=
\begin{pmatrix}
1 & & & \\
& q & & \\
& & \ddots & \\
& & &q^{d-1}
\end{pmatrix},
\label{Iqsigma}
\end{equation}
the following parameterization provides solutions to \eqref{rttrelation}
\begin{gather}
a = x \iq, \quad b = y \sigma, \quad c = z \sigma, \nonumber\\
d = a^{-1}+b a^{-1} c = \frac{1}{x}\text{I}_{q^{-1}}+\frac{yz}{x} \sigma\text{I}_{q^{-1}}\sigma,
\label{canonicalform}
\end{gather}
with $x$, $y$ and $z$ free complex scalar parameters and $x \neq 0$. This parametrization turns $T$ into a four-index tensor that we shall denote as the set of $d\times d$ matrices $T^{ij}$, where $i,j = 0,1$ denote the 2-dimensional physical indices, and the dependence on $x,y,z$ is implicit. These tensors allow us to define the three-parameter family of MPOs with periodic boundary conditions on $L$ sites as
\begin{align}
\mathcal{T}(x,y,z)&=\!\!\!\!\sum_{\{i\}\{j\}=0}^1\!\!\!\! \Tr\left(T^{i_1j_1}\dots\nonumber T^{i_Lj_L}\right)\ket{i_1\dots i_L}\bra{j_1 \dots j_L}\\
&=\quad
\begin{tikzpicture}[scale=1,baseline={([yshift=-0.5ex]current bounding box.center)}]
\draw [red, line width=.8pt] (-1.5,0) -- (0.5,0); 
\draw [red, line width=.8pt] (1.5,0) -- (0.5,0); 
\draw [dotted,red, line width=.8pt] (1.5,0) -- (2,0); 
\draw [dotted,red, line width=.8pt] (-1.5,0) -- (-2,0); 
\draw [black, line width=.8pt] (-1,-0.5) -- (-1,0.5); 
\draw [black, line width=.8pt] (1,-0.5) -- (1,0.5); 
\draw [black, line width=.8pt] (0,-0.5) -- (0,0.5); 
\draw[black,fill=white] (-1,0) circle (1.5ex);
\draw[black,fill=white] (1,0) circle (1.5ex);
\draw[black,fill=white] (0,0) circle (1.5ex);
\draw (1,0) node {\small$T$};
\draw (-1,0) node {\small$T$};
\draw (0,0) node {\small$T$};
\end{tikzpicture}
\label{mpodef},
\end{align}
where from now on we use the diagrammatic notation to depict tensor contractions. Let us consider the following transformations on the virtual degrees of freedom:
\begin{align}
T(x,y,z)^{ij} &\rightarrow \sigma T(x,y,z)^{ij} \sigma^{-1} = T(qx,y,z)^{ij},\nonumber\\
T(x,y,z)^{ij} &\rightarrow I_q T(x,y,z)^{ij}{I_q}^{-1} = T(x,q^{-1}y,q^{-1}z)^{ij}.
\end{align}
These transformations leave the periodic boundary condition MPO  $\mathcal{T}(x, y, z)$ invariant. This implies that the parameterization of the MPOs $\mathcal{T}$ in terms of $(x,y,z)$ is not one-to-one, since among the $d^3$ MPOs $\mathcal{T}(q^ix,q^jy,q^kz)$ with $i,j,k = 1,\ldots,d$, only $d$ are distinct. Therefore we change our parameterization to
$(\tilde{x},\tilde{y},\tilde{z}) \equiv (x^d, y^d, z^d)$. This identifies the labels $g$ with the central characters of $\mathcal{O}_q(SL_2)$ at a root of unity \cite{ParshallWang1991,DeConciniLyubashenko1994}. We therefore define
$d$ generically distinct MPOs
\begin{align}
D_k(\tilde{x},\tilde{y},\tilde{z})^{ij} &\equiv T(\tilde{x}^{1/d},\tilde{y}^{1/d},q^k\tilde{z}^{1/d})^{ij}, \nonumber\\
\mathcal{D}_k(\tilde{x},\tilde{y},\tilde{z}) &\equiv \mathcal{T}(\tilde{x}^{1/d},\tilde{y}^{1/d},q^k\tilde{z}^{1/d}),
\end{align}
where the label $k$ is arbitrary, but has to be chosen in a consistent way. From now on, we will use the shorthand notation $g$ for the objects $(\tilde{x},\tilde{y},\tilde{z})$.

\paragraph{Fusion of two MPOs:} The multiplication of two such MPOs turns out to form a closed MPO algebra \cite{bultinck2017anyons} with structure constants that are independent of the number of sites.  The product of two MPOs is generated by the operators
\begin{equation}
    \begin{tikzpicture}[scale=1,baseline={([yshift=-0.5ex]current bounding box.center)}]
        \draw [red, line width=.8pt] (-.8,0) -- (1.2,0);
        \draw [red, line width=.8pt] (-.8,1) -- (1.2,1); 
        \draw [black, line width=.8pt] (0,-0.5) -- (0,1.5); 
        \draw [black,fill=white] (0,0) circle (1.5ex);
        \draw [black,fill=white] (0,1) circle (1.5ex);
        \draw (0,0) node {\small$D$};
        \draw (0,1) node {\small$D$};
        \draw [red] (0.7,1.2) node {\footnotesize$(g_1,k_1)$};
        \draw [red] (0.7,0.2) node {\footnotesize$(g_2,k_2)$};
        \draw (-1.1,1) node {\footnotesize$n$};
        \draw (1.5,1) node {\footnotesize$n'$};
        \draw (-1.1,0) node {\footnotesize$m$};
        \draw (1.5,0) node {\footnotesize$m'$};
        \draw (0,1.7) node {\footnotesize$i$};
        \draw (0,-0.7) node {\footnotesize$j$};
    \end{tikzpicture} = 
    [M^{ij}(g_1, k_1, g_2, k_2)]_{dn+m}^{dn' + m'}
\label{eq:MPOVerticallyFused}
\end{equation}
and given by a $d^2 \times d^2$ square matrix with row indices $(n, m)$ and column indices $(n', m')$. We use the notation $M^{00} = A$, $M^{01} = B$, $M^{10} = C$, $M^{11} = D$, where we omitted the dependence on the representation labels $(g_i, k_i)$. Note that $A,B,C,D$ also satisfy the RTT relations \eqref{rttrelation}. 

For the sake of the argument, let us restrict ourselves to the case $d=3$, although the general case (appendix) is very similar. Due to  \eqref{rttrelation}, we have that $Q_1=B^2C$ and $Q_2=BC^2$, together with the identity, span the center of the algebra. Furthermore, it holds that $B^3=f_1\mathds{1}$ and $C^3=f_2\mathds{1}$ are both proportional to the identity with $f_i$ scalar functions of all variables involved (a complication will arise when they are nilpotent; we will come back to this case later). It follows that $Q_1^2=f_1Q_2$, $Q_2^2=f_2Q_1$ and $Q_1Q_2=Q_2Q_1=f_1f_2\mathds{1}$. Making use of these algebraic relations, we can construct three orthogonal projectors $P_iP_j=\delta_{ij}P_i$:
\begin{equation}
    P_i=\frac{1}{3}\left(\mathds{1}+\frac{q^i}{f_1^{2/3}f_2^{1/3}}Q_1+\frac{q^{-i}}{f_1^{1/3}f_2^{2/3}}Q_2\right)
\end{equation}
As these three projectors are built from elements in the commutant of the algebra, their images are invariant subspaces. By some cumbersome but straightforward algebra (see Appendix B) , it can be shown that for generic $g_1,g_2$ and any odd $d$
\begin{align}
\mathcal{D}_{k_1}(g_1)\mathcal{D}_{k_2}(g_2)=\sum_{k_{12}=0}^{d-1}\mathcal{D}_{k_{12}}(g_{12})\label{simplefusion}
\end{align}
where $g_i\equiv(\tilde{x}_i,\tilde{y}_i,\tilde{z}_i)$ label elements of the Lie group $SL(2)$, $\tilde{x}_i\tilde{t}_i-\tilde{y}_i\tilde{z}_i=1$ and 
\begin{align}
\left( \begin{array}{cc} \td{x}_{12} & \td{y}_{12} \\
\td{z}_{12} & \td{t}_{12} \end{array} \right)=
\left( \begin{array}{cc} \td{x}_1 & \td{y}_1 \\
\td{z}_1 & \td{t}_1 \end{array} \right)
\left( \begin{array}{cc} \td{x}_2 & \td{y}_2 \\
\td{z}_2& \td{t}_2 \end{array} \right)
\label{su2rule}.
\end{align}
Hence $g_{12}=g_1\cdot g_2$ with $\cdot$ the product of the Lie group $SL(2)$, demonstrating that  we indeed recover the $SL(2)$ group structure starting from the matrix quantum group structure. Note that the defining scalars of the MPO are obtained by taking $d$-th roots of these elements;  as we will show later, this is related to the fact that the corresponding raising operator in this MPO representation of the Lie algebra of $SL(2)$ flips $d$ spins. It is now natural to define non-injective MPO tensors $M(g)^{ij}$ (see Eq. \eqref{eq:MPOVerticallyFused}) of bond dimension $d^2$ and their corresponding MPOs $\mathcal{D}(g)$ as
\begin{equation}
M(g)^{ij} = \bigoplus_{k=0}^{d-1} D_k(g)^{ij}, \quad \mathcal{D}(g)=\frac{1}{d^2}\sum_{k=0}^{d-1}\mathcal{D}_k(g)
\label{superposition}.
\end{equation}
These MPOs form a representation of $SL(2)$:
\begin{equation}
\mathcal{D}(g_1)\mathcal{D}(g_2)=\mathcal{D}(g_{12}).
\label{fullfusion}
\end{equation}

The classical group $SL(2)$ emerges inside the matrix quantum group, and the multiplication table is exactly the same as the one derived in the context of Lusztig's quantum Frobenius map \cite{ParshallWang1991,DeConciniLyubashenko1994}! Our derivation and MPO representation is very different, however. The quantum Frobenius map is induced by the map
\begin{equation}
    \begin{pmatrix} a & b \\ c & d\end{pmatrix}\ \longrightarrow\ \begin{pmatrix} a^d & b^d \\ c^d & d^d\end{pmatrix}\equiv\begin{pmatrix}\tilde{x} & \tilde{y}\\ \tilde{z} & \tilde{t}\end{pmatrix}\otimes\mathbb{I},
\end{equation}
which is $d$-to-one on gauge equivalence classes and whose image consists of central elements. That it intertwines the coproduct is what makes it a Hopf algebra map; for instance,
\begin{equation}
    A^d=\left(a_1\otimes a_2+b_1\otimes c_2\right)^{d}=\left(\tilde{x}_1\tilde{x}_2+\tilde{y}_1\tilde{z}_2\right)\mathbb{I}.
\end{equation}
Our MPO construction realizes the same group law by a very different route, one involving all monomials of degree 
$L$ in  $a,b,c,d$; the multiplication table nevertheless comes out identical. As a bonus, we obtain recoupling coefficients on the virtual level that have no counterpart in the algebraic construction, and the fusion reveals a very interesting  non-semisimple structure.

\paragraph{Special cases}
The fusion rules above indeed break down in a number of special cases. This is related to the existence and diagonalisability of non-trivial central idempotents in the algebra spanned by the fused tensors, which in turn depends on whether the central elements $Q_1=B^2C$ and $Q_2=BC^2$ are nilpotent. A detailed analysis of when special cases occur can be found in Appendix A.

If $B$ and $C$ are both non-zero and at least one of $B, C$ is nilpotent, the fused representation is reducible but indecomposable -- a hallmark of quantum group representations at roots of unity. In bases adapted to invariant subspaces, Jordan blocks appear. For instance, if the product of two non-diagonal matrices results in a diagonal matrix, both $B$ and $C$ are nilpotent. If they are non-zero, the corresponding MPO-tensors are not semisimple and equivalent to the following indecomposable $d^2\times d^2$ matrices:
\begin{equation}
\begin{aligned}
A_k
&=
q^{k}\Lambda_A^{1/d}
\left(I_q\otimes I_q\right)
\\[0.6em]
B
&=J_d\otimes \mathds{1}_d \\[0.8em]
C
&=\mathds{1}_d\otimes J_d\\[0.8em]
D_k&=A_k^{-1}+BA_k^{-1}C,
\end{aligned}
\label{indid}
\end{equation}
where $\Lambda_A^{1/d}$ is some reference root of $\Lambda_A = x_1^d x_2^d + y_1^d z_2^d$, and $k$ depends on which eigenvalue of $A$ is the eigenvalue of the unique $A$-eigenvector in $\ker B \cap \ker C$. $I_q$ and $\sigma$ are the matrices \eqref{Iqsigma}, $J_d$ is the $d\times d$  upper triangular Jordan matrix,  while $\mathds{1}_d$ is a $d\times d$ identity matrix. $A_k, D_k$ are dependent on the $k$ label of the fused object. For fixed $k_1, k_2$, a single $k$ is selected; as the input labels vary, each possible $k$ occurs exactly $d$ times.  For details and other similar cases (e.g. when exactly one of $B, C$ is nilpotent), see Appendix E.

Let us now consider the case where  $A$ or $D$ are nilpotent. Then the fused representation is still decomposable, but the individual blocks in the direct sum are nilpotent themselves, thus admitting a basis where $A$ or $D$ is a direct sum of Jordan matrices. For instance, if $A$ is nilpotent, it becomes non-diagonalisable, and acquires a non-zero kernel.  There exists a basis in which the resulting MPO entries are of the form:
\begin{align}
    A &= \mathds{1} \otimes J \\
    B &= \frac{q^{-1}y_1}{x_2} I_q \otimes I_{q^{-1}} \\
    C &= -\frac{x_2}{y_1} I_{q^{-1}} \otimes I_{q^{-1}} \\
    D &= \mathds{1} \otimes \tilde{D}, \; \\
    \tilde{D}_{i, i-1} &= 1-q^{-2i} \; \text{for} \; i > 0, \; \tilde{D}_{0, d-1} = \frac{\Lambda_D}{d} 
\end{align}
where $\Lambda_{D} = \frac{1+y_1^d z_1^d}{x_1^d x_2^d} - \frac{y_2^d}{y_1^d}$, i.e. the unique eigenvalue of $D^d$. Of course, as $\mathds{1}$ and $I_q$ are diagonal, those are block-diagonal forms. The case with $D$ nilpotent is analogous, with $A$ and $D$ switching places --- $A$ acts like a cyclic shift on the $D$-filtrated basis, i.e. union of bases of $\ker D^i / \ker D^{i-1}$, with $i \in \{1, ..., d\}$, which simultaneously diagonalise $B$ and $C$.

We checked that in all cases the algebra is still closed: multiplying two MPOs of the form \eqref{indid}  yields an MPO of bond dimension $81$ that is reducible to the same indecomposable $9\times 9$ blocks \eqref{indid}.

\paragraph{Bimodule categories and pentagon equations:} The tensors constructed above fit naturally in the categorical language of MPO symmetries \cite{LFHSV}, in which local MPO tensors play the role of the associators ($^2F$) of a bimodule category and the fusion tensors computed in Appendix B are horizontal zipper tensors ($^1F$): generalized Clebsch-Gordan coefficients that project the product of the irreps $(g_1,k_1)$ and $(g_2,k_2)$ onto the irrep $(g_1g_2,k_{12})$, satisfying the zipper or \emph{pulling-through} equation. Note that for the special cases, the indecomposable objects will contain Jordan block structures with indecomposable but reducible $d^2$-dimensional blocks (still labelled by an extra index $k$). Requiring associativity of this fusion up to isomorphism yields pentagon-type equations whose solutions define the associators $^0F$; all internal multiplicity spaces are $0$- or $1$-dimensional, and objects labelled by elements of $SL(2)$ other than the product $g_1g_2$ do not appear in the decomposition. For the special cases however, there is no guarantee anymore that the associators are scalar. The corresponding vertical zipper tensors ($^3F$) are the $q$-deformed Clebsch-Gordan coefficients, with recoupling governed by the $q$-deformed $6j$ symbols ($^4F$) of $U_q(sl_2)$ \cite{AS10}; this vertical problem is well understood since the  work of Pasquier and Saleur \cite{PS90}, which showed that the relevant modules are reducible but indecomposable --- our MPO results can be read as a dual manifestation of that property.

\paragraph{Relation to the generators:} Let us go back to the  MPO representations. As 
they form a representation of the group $SL(2)$, we can define generators of the Lie algebra by taking infinitesimal deformations around the identity element $(\td{x},\td{y},\td{z})=(1,0,0)$.  The corresponding ``identity'' MPO, defined  on $L$ sites,  is not equal to the identity but is rather a projector onto the subspace of all states for which the magnetization in the $S_z$-direction is a multiple of $d$; this follows from the fact that 
\begin{equation}
\sum_{k=0}^{d-1}q^{kS_z}=\left\{\begin{array}{rcl}
	d & \mbox{if} & S_z=0 \text{ mod } d \\ 0 & \mbox{else} &
\end{array}\right.
\end{equation}
We now define the following operators:
\begin{equation}
J_i=(\partial_i\mathcal{D}(g))|_{g=(1,0,0)},\quad i=1,2,3.
\end{equation}
As expected, their commutation relations are dictated by the Lie algebra:
\begin{equation}
\left[J_i,J_j\right]=i\epsilon_{ijk}J_k
\end{equation}

For the case $d=3$, we write them out explicitly as 
\begin{align}
    &J_1=\frac{1}{2d^2}\sum_{r=0}^2\sum_{m=0}^2\sum_{1\leq i_1<i_2<i_3\leq L}\left(q^{m\sigma_z}\right)^{\otimes(i_1-1)}\nonumber\\&\otimes \sigma_{x,q^r}\otimes \left(q^{(m-1)\sigma_z}\right)^{\otimes(i_2-i_1-1)}\otimes\sigma_{x,q^r}\nonumber\\&\otimes \left(q^{(m-2)\sigma_z}\right)^{\otimes(i_3-i_2-1)}\otimes\sigma_{x,q^r}\otimes
    \left(q^{m\sigma_z}\right)^{\otimes(L-i_3)}.
\end{align}
where $\sigma_{x,q^r}=\begin{pmatrix} 
0 & 1\\
q^r & 0
\end{pmatrix}$. 
$J_2$ is obtained by replacing $\sigma_{x,q^r}$ by $\sigma_{y,q^r}=\begin{pmatrix} 0 & -i\\
q^ri & 0
\end{pmatrix}$, and $J_3= -\frac{1}{2d} P_d S_z$ where $P_d = \frac{1}{d}\sum_{i} q^{i S_{z}}$ and $S_z := \sum_i \sigma_z(i)$, i.e. $J_3$ is the usual total magnetization operator projected on the subspace with magnetization $0$ mod $d$. The operators $J_1$ and $J_2$ are hence built from $d$ raising or lowering operators; as such, we recover an expression that is not very dissimilar to the one found in \cite{deguchi2001sl2}, although here there is an extra sum over the indices $r$ and $m$.   In \cite{deguchi2001sl2}, the generators of the loop algebra $SL_2$ were obtained by taking a particular limit of powers of the quantum group generators of $U_q(\mathfrak{sl}(2))$ \cite{PS90}. In contrast to their work, the algebra that we obtain is finite dimensional, which follows from the triple sum. Our work therefore  makes an explicit connection between the quantum group as a deformed Lie algebra and its group-like elements, analogous to the exponential map from the Lie algebra to the Lie group~\footnote{Note however that our MPOs are not simply the exponential map applied to these operators, but rather the exponential map followed by a projection on the subspace with a charge proportional to $d$.}. From the conceptual point of view, this might well be one of the most interesting results of this paper: MPO algebras provide the framework for defining exponential maps of quantum group generators.

\paragraph{Summary, applications and outlook} Generalized symmetries and their description in terms of fusion categories have recently taken the theoretical physics world by storm as they provide a very general and powerful framework for describing anomalies in quantum many-body systems. However, an extension to generalized symmetries described by continuous parameters has been lacking. In this paper, we provide first evidence that quantum groups and their MPO representations fill this void. 

From a technical point of view, we have established a general framework for defining the ``exponential map'' of a quantum group algebra. The resulting MPO algebra turns out to have a host of interesting properties, such as containing reducible but indecomposable modules. Altogether, this work solidifies the role of MPO symmetries as convenient representations of novel generalised symmetries, extending naturally beyond the finite, semi-simple case of fusion-categorical symmetries. 

Many interesting models exhibit quantum group symmetries. The best known is the $XXZ$ spin chain \cite{PS90}, the Hamiltonian $H = \sum_i h_i$ of which can be expressed in terms of the above mentioned Clebsch-Gordan coefficients as
\begin{equation}
h_i =
\begin{tikzpicture}[scale=0.6,baseline={([yshift=-0.5ex]current bounding box.center)}]
\draw [black, line width=.8pt] (0,0.5) -- (0,1); 
\draw [black, line width=.8pt] (1,0.5) -- (1,1); 
\draw (1,2) node {\small$0$};
\draw (-0.6,0.8) node {\small$1/2$};
\draw (1.5,0.8) node {\small$1/2$};
\draw (-0.6,3.3) node {\small$1/2$};
\draw (1.5,3.3) node {\small$1/2$};
\draw (0,0) node {\small$i$};
\draw (1,0) node {\small$i\!\!+\!\!1$};
\draw[black, line width=.8pt](1,1) arc (0:180:0.5);
\draw [black, line width=.8pt] (0.5,1.5) -- (0.5,2.5); 
\draw[black, line width=.8pt](0,3) arc (180:360:0.5);
\draw [black, line width=.8pt] (0,3.5) -- (0,3); 
\draw [black, line width=.8pt] (1,3.5) -- (1,3);
\end{tikzpicture}.
\end{equation}
The vertical zipper (pulling-through) condition then implies that the MPOs we constructed above are indeed symmetries of this model. At criticality, these models are described by logarithmic conformal field theories \cite{fuchs2019full}, which are characterized by the presence of reducible but indecomposable representations of the Virasoro algebra. In general, the presence of MPO symmetries signals the fact that one can construct excitations in systems with those MPO symmetries by considering the tube algebra associated to these MPOs \cite{bultinck2017anyons,marien2017condensation}. Similarly, these MPO symmetries allow one to define twisted boundary conditions, which can also be understood via the action of the tube algebra \cite{aasen2016topological,vanhove2018mapping,aasen2020topological,vanhove2022topological}. It would also be very interesting to investigate the consequences of our formalism for describing topological defects --- and the ensuing conversion of wavepackets into string-like objects~\cite{ueda2025perfect}. For integrable lattice models exhibiting these MPO symmetries on open boundary conditions, the action of an MPO symmetry on an integrable boundary condition will generate a new integrable boundary condition, a procedure that is well known in the integrable lattice model literature \cite{behrend1998integrable}. Importantly however, these MPO symmetries are not restricted to integrable lattice models, although we expect their presence, supplemented with certain conserved current conditions, to imply integrability \cite{fendley2021integrability}. If true, satisfying the pulling-through condition of some quantum group MPO symmetry and checking a current conservation law could yield a convenient check of integrability.

Finally, let us briefly comment on the even $d$ case which leads to an even richer structure. Here, the $d/2$-th power controls the fusion rules, as we have $A^{d/2}=a_1^{d/2}\otimes a_2^{d/2}+b_1^{d/2}\otimes c_2^{d/2}$ and  $a^{d/2}d^{d/2}-(-1)^{d/2}b^{d/2}c^{d/2}=1$. However, the $d/2$-th powers cease to be central and anticommute with the generators, so that they span a $\mathbb{Z}_{2}$-graded rather than an ordinary centre. The MPO then vanishes identically on odd rings, while on even rings the product of two MPOs is a direct sum of two, with labels $g_{1}g_{2}$ and $\pi(g_{1})g_{2}$, where $\pi$ is the parity automorphism.  The group realised at even roots of unity is thus not $SL(2)$ itself but its $\mathbb{Z}_{2}$ orbifold by fermion parity, creating a direct link to graded tensor networks~\cite{Bultinck_2017}. This will be reported in a subsequent publication.

\paragraph{Acknowledgements:} We are  grateful to Jacob Bridgeman for interesting discussions. L.L. is supported by an EPSRC Postdoctoral Fellowship (grant No. EP/Y020456/1). FV acknowledges funding from the UKRI grant EP/Z003342/1, BOFGOA (Grant No. BOF23/GOA/021) and IBOF (Grant No.~IBOF23/064).

\bibliography{bib}

@article{Lusztig1989,
  author  = {Lusztig, George},
  title   = {Modular representations and quantum groups},
  journal = {Contemp. Math.},
  volume  = {82},
  pages   = {59--77},
  year    = {1989},
}

@article{Lusztig1990jams,
  author  = {Lusztig, George},
  title   = {Finite dimensional {H}opf algebras arising from quantized
             universal enveloping algebras},
  journal = {J. Amer. Math. Soc.},
  volume  = {3},
  pages   = {257--296},
  year    = {1990},
}

@book{Lusztig1993book,
  author    = {Lusztig, George},
  title     = {Introduction to Quantum Groups},
  publisher = {Birkh\"auser},
  address   = {Boston},
  year      = {1993},
}

@book{ParshallWang1991,
  author    = {Parshall, Brian and Wang, Jian-pan},
  title     = {Quantum Linear Groups},
  series    = {Mem. Amer. Math. Soc.},
  volume    = {439},
  publisher = {American Mathematical Society},
  year      = {1991},
}

@article{DeConciniLyubashenko1994,
  author  = {De Concini, Corrado and Lyubashenko, Volodymyr},
  title   = {Quantum function algebra at roots of 1},
  journal = {Adv. Math.},
  volume  = {108},
  pages   = {205--262},
  year    = {1994},
}

@misc{rubio2026,
      title={The local characterization of global tensor network eigenstates}, 
      author={José Garre Rubio and András Molnár and Norbert Schuch and Frank Verstraete},
      year={2026},
      eprint={2603.28349},
      archivePrefix={arXiv},
      primaryClass={quant-ph},
      url={https://arxiv.org/abs/2603.28349}, 
}

@article{Ilievski_2016,
   title={Quasilocal charges in integrable lattice systems},
   volume={2016},
   ISSN={1742-5468},
   url={http://dx.doi.org/10.1088/1742-5468/2016/06/064008},
   DOI={10.1088/1742-5468/2016/06/064008},
   number={6},
   journal={Journal of Statistical Mechanics: Theory and Experiment},
   publisher={IOP Publishing},
   author={Ilievski, Enej and Medenjak, Marko and Prosen, Tomaž and Zadnik, Lenart},
   year={2016},
   month=jun, 
   pages={064008} 
}

@article{PROSEN20141177,
title = {Quasilocal conservation laws in XXZ spin-1/2 chains: Open, periodic and twisted boundary conditions},
journal = {Nuclear Physics B},
volume = {886},
pages = {1177-1198},
year = {2014},
issn = {0550-3213},
doi = {https://doi.org/10.1016/j.nuclphysb.2014.07.024},
url = {https://www.sciencedirect.com/science/article/pii/S0550321314002430},
author = {Tomaž Prosen}
}

@article{Pereira_2014,
   title={Exactly conserved quasilocal operators for the XXZ spin chain},
   volume={2014},
   ISSN={1742-5468},
   url={http://dx.doi.org/10.1088/1742-5468/2014/09/P09037},
   DOI={10.1088/1742-5468/2014/09/p09037},
   number={9},
   journal={Journal of Statistical Mechanics: Theory and Experiment},
   publisher={IOP Publishing},
   author={Pereira, R G and Pasquier, V and Sirker, J and Affleck, I},
   year={2014},
   pages={P09037} }

@article{ueda2025perfect,
  author   = {Ueda, Atsushi and Vander Linden, Vic and De Vos, Boris and Lootens, Laurens and Haegeman, Jutho and Fendley, Paul and Verstraete, Frank},
  title    = {Perfect particle transmission through duality defects},
  journal  = {Nature Physics},
  year     = {2026},
  month    = aug,
  doi      = {10.1038/s41567-026-03390-5},
  url      = {https://doi.org/10.1038/s41567-026-03390-5},
  issn     = {1745-2481},
  eprint        = {2510.26780},
}

@article{Polchinski:fermion_rotor,
    author = "Polchinski, Joseph",
    title = "{Monopole Catalysis: The Fermion Rotor System}",
    reportNumber = "HUTP-84-A005",
    doi = "10.1016/0550-3213(84)90398-5",
    journal = "Nucl. Phys. B",
    volume = "242",
    pages = "345--363",
    year = "1984"
}

@article{vanBeest:monopole,
  title={Monopoles, scattering, and generalized symmetries},
  author={Marieke van Beest and Philip Smith and Diego Delmastro and Zohar Komargodski and David Tong},
  journal={Journal of High Energy Physics},
  volume = {03},
  pages = {014},
    year = {2025},
  url={https://api.semanticscholar.org/CorpusID:259145333},
  eprint = "2306.07318",
    archivePrefix = {arXiv},
    primaryClass = {hep-th},
}

@article{Loladze:2025jsq,
  title={Dynamics of the fermion-rotor system},
  author={Loladze, Vazha and Okui, Takemichi and Tong, David},
  journal={Journal of High Energy Physics},
  volume={2026},
  number={1},
  pages={52},
  year={2026},
  publisher={Springer}
}

@article{Callan:1982ah,
    author = "Callan, Jr., Curtis G.",
    title = "{Disappearing Dyons}",
    reportNumber = "Print-82-0006 (PRINCETON)",
    doi = "10.1103/PhysRevD.25.2141",
    journal = "Phys. Rev. D",
    volume = "25",
    pages = "2141",
    year = "1982"
}

@article{Rubakov:1982fp,
    author = "Rubakov, V. A.",
    title = "{Adler-Bell-Jackiw Anomaly and Fermion Number Breaking in the Presence of a Magnetic Monopole}",
    doi = "10.1016/0550-3213(82)90034-7",
    journal = "Nucl. Phys. B",
    volume = "203",
    pages = "311--348",
    year = "1982"
}

@article{Lusztig1990,
  author   = {Lusztig, George},
  title    = {Quantum Groups at Roots of 1},
  journal  = {Geometriae Dedicata},
  year     = {1990},
  volume   = {35},
  number   = {1},
  pages    = {89--113},
  doi      = {10.1007/BF00147341},
  url      = {https://doi.org/10.1007/BF00147341},
  issn     = {1572-9168}
}

@article{de1992quantum,
  title={Quantum coadjoint action},
  author={De Concini, Corrado and Kac, Victor G and Procesi, Claudio},
  journal={Journal of the American Mathematical Society},
  volume={5},
  number={1},
  pages={151--189},
  year={1992},
  publisher={JSTOR}
}

@article{Brandino_2015,
   title={Glimmers of a Quantum KAM Theorem: Insights from Quantum Quenches in One-Dimensional Bose Gases},
   volume={5},
   ISSN={2160-3308},
   url={http://dx.doi.org/10.1103/PhysRevX.5.041043},
   DOI={10.1103/physrevx.5.041043},
   number={4},
   journal={Physical Review X},
   publisher={American Physical Society (APS)},
   author={Brandino, G. P. and Caux, J.-S. and Konik, R. M.},
   year={2015},
   month=dec
}

@article{Miao_2021,
   title={On the Q operator and the spectrum of the XXZ model at root of unity},
   volume={11},
   ISSN={2542-4653},
   url={http://dx.doi.org/10.21468/SciPostPhys.11.3.067},
   DOI={10.21468/scipostphys.11.3.067},
   number={3},
   journal={SciPost Physics},
   publisher={Stichting SciPost},
   author={Miao, Yuan and Lamers, Jules and Pasquier, Vincent},
   year={2021},
   month=sep
}

@article{Fabricius_2001, 
   volume={103},
   ISSN={0022-4715},
   url={http://dx.doi.org/10.1023/A:1010380116927},
   DOI={10.1023/a:1010380116927},
   number={5/6},
   journal={Journal of Statistical Physics},
   publisher={Springer Science and Business Media LLC},
   author={Fabricius, Klaus and McCoy, Barry M.},
   year={2001},
   pages={647–678}
}

@article{Fabricius:2000em,
    author = "Fabricius, Klaus and McCoy, Barry M.",
    title = "{Completing Bethe's equations at roots of unity}",
    eprint = "cond-mat/0012501",
    archivePrefix = "arXiv",
    reportNumber = "YITP-SB-00-89",
    doi = "10.1023/A:1010372504158",
    journal = "J. Statist. Phys.",
    volume = "104",
    pages = "573--587",
    year = "2001"
}

@article{deguchi2001sl2,
  title={The sl2 loop algebra symmetry of the six-vertex model at roots of unity},
  author={Deguchi, Tetsuo and Fabricius, Klaus and McCoy, Barry M},
  journal={Journal of Statistical Physics},
  volume={102},
  number={3},
  pages={701--736},
  year={2001},
  publisher={Springer}
}

@article{buican2017anyonic,
  title={Anyonic chains, topological defects, and conformal field theory},
  author={Buican, Matthew and Gromov, Andrey},
  journal={Communications in Mathematical Physics},
  volume={356},
  number={3},
  pages={1017--1056},
  year={2017},
  publisher={Springer}
}

@article{cirac2021matrix,
  title={Matrix product states and projected entangled pair states: Concepts, symmetries, theorems},
  author={Cirac, J Ignacio and Perez-Garcia, David and Schuch, Norbert and Verstraete, Frank},
  journal={Reviews of Modern Physics},
  volume={93},
  number={4},
  pages={045003},
  year={2021},
  publisher={APS}
}

@Article{LFHSV,
	title={{Matrix product operator symmetries and intertwiners in string-nets with  domain walls}},
	author={Laurens Lootens and Jürgen Fuchs and Jutho Haegeman and Christoph Schweigert and Frank Verstraete},
	journal={SciPost Phys.},
	volume={10},
	issue={3},
	pages={53},
	year={2021},
	publisher={SciPost},
	doi={10.21468/SciPostPhys.10.3.053},
	url={https://scipost.org/10.21468/SciPostPhys.10.3.053},
}

@article{schuch2010peps,
  title={PEPS as ground states: Degeneracy and topology},
  author={Schuch, Norbert and Cirac, Ignacio and P{\'e}rez-Garc{\'\i}a, David},
  journal={Annals of Physics},
  volume={325},
  number={10},
  pages={2153--2192},
  year={2010},
  publisher={Elsevier}
}

@article{vanhove2018mapping,
  title={Mapping topological to conformal field theories through strange correlators},
  author={Vanhove, Robijn and Bal, Matthias and Williamson, Dominic J and Bultinck, Nick and Haegeman, Jutho and Verstraete, Frank},
  journal={Physical review letters},
  volume={121},
  number={17},
  pages={177203},
  year={2018},
  publisher={APS}
}

@article{aasen2016topological,
  title={Topological defects on the lattice: I. The Ising model},
  author={Aasen, David and Mong, Roger SK and Fendley, Paul},
  journal={Journal of Physics A: Mathematical and Theoretical},
  volume={49},
  number={35},
  pages={354001},
  year={2016},
  publisher={IOP Publishing}
}

@article{gils2013anyonic,
  title={Anyonic quantum spin chains: Spin-1 generalizations and topological stability},
  author={Gils, Charlotte and Ardonne, Eddy and Trebst, Simon and Huse, David A and Ludwig, Andreas WW and Troyer, Matthias and Wang, Zhenghan},
  journal={Physical Review B},
  volume={87},
  number={23},
  pages={235120},
  year={2013},
  publisher={APS}
}

@article{pirvu2010matrix,
  title={Matrix product operator representations},
  author={Pirvu, Bogdan and Murg, Valentin and Cirac, J Ignacio and Verstraete, Frank},
  journal={New Journal of Physics},
  volume={12},
  number={2},
  pages={025012},
  year={2010},
  publisher={IOP Publishing}
}

@article{bultinck2017anyons,
  title={Anyons and matrix product operator algebras},
  author={Bultinck, Nick and Mari{\"e}n, Michael and Williamson, Dominic J and {\c{S}}ahino{\u{g}}lu, Mehmet B and Haegeman, Jutho and Verstraete, Frank},
  journal={Annals of physics},
  volume={378},
  pages={183--233},
  year={2017},
  publisher={Elsevier}
}

@article{PS90,
title = {Common structures between finite systems and conformal field theories through quantum groups},
journal = {Nuclear Physics B},
volume = {330},
number = {2},
pages = {523-556},
year = {1990},
issn = {0550-3213},
doi = {https://doi.org/10.1016/0550-3213(90)90122-T},
url = {https://www.sciencedirect.com/science/article/pii/055032139090122T},
author = {V. Pasquier and H. Saleur}
}

@book{Biedenharn,
author = {Biedenharn, L C and Lohe, M A},
title = {Quantum Group Symmetry and Q-Tensor Algebras},
publisher = {World Scientific},
year = {1995},
doi = {10.1142/2815},
address = {},
edition   = {}
}

@article{AS10,
	doi = {10.1088/1751-8113/43/39/395205},
	url = {https://doi.org/10.1088/1751-8113/43/39/395205},
	year = 2010,
	month = {aug},
	publisher = {{IOP} Publishing},
	volume = {43},
	number = {39},
	pages = {395205},
	author = {Eddy Ardonne and Joost Slingerland},
	title = {Clebsch{\textendash}Gordan and 6j-coefficients for rank 2 quantum groups},
	journal = {Journal of Physics A: Mathematical and Theoretical}
}

@article{LDOV21,
  title         = {Dualities in One-Dimensional Quantum Lattice Models:
                   Symmetric Hamiltonians and Matrix Product Operator
                   Intertwiners},
  author        = {Lootens, Laurens and Delcamp, Clement and Ortiz, Gerardo
                   and Verstraete, Frank},
  journal       = {PRX Quantum},
  volume        = {4},
  pages         = {020357},
  year          = {2023},
  doi           = {10.1103/PRXQuantum.4.020357},
  eprint        = {2112.09091},
  archivePrefix = {arXiv},
  primaryClass  = {quant-ph},
}

@article{Bultinck_2017,
	doi = {10.1088/1751-8121/aa99cc},
	url = {https://doi.org/10.1088/1751-8121/aa99cc},
	year = 2017,
	month = {dec},
	publisher = {{IOP} Publishing},
	volume = {51},
	number = {2},
	pages = {025202},
	author = {Nick Bultinck and Dominic J Williamson and Jutho Haegeman and Frank Verstraete},
	title = {Fermionic projected entangled-pair states and topological phases},
	journal = {Journal of Physics A: Mathematical and Theoretical}
}

@article{VJS11,
  title={Indecomposability parameters in chiral logarithmic conformal field theory},
  author={Vasseur, Romain and Jacobsen, Jesper Lykke and Saleur, Hubert},
  journal={Nuclear Physics B},
  volume={851},
  number={2},
  pages={314--345},
  year={2011},
  publisher={Elsevier}
}

@article{BGJST20,
  title={Topological defects in periodic RSOS models and anyonic chains},
  author={Bellet{\^e}te, J and Gainutdinov, AM and Jacobsen, JL and Saleur, H and Tavares, TS},
  journal={arXiv preprint arXiv:2003.11293},
  year={2020}
}

@book{etingof2016tensor,
  title={Tensor categories},
  author={Etingof, Pavel and Gelaki, Shlomo and Nikshych, Dmitri and Ostrik, Victor},
  volume={205},
  year={2016},
  publisher={American Mathematical Soc.}
}

@incollection{faddeev1988quantization,
  title={Quantization of Lie groups and Lie algebras},
  author={Faddeev, Ludwig D and Reshetikhin, N Yu and Takhtajan, Leon A},
  booktitle={Algebraic analysis},
  pages={129--139},
  year={1988},
  publisher={Elsevier}
}

@article{woronowicz1987compact,
  title={Compact matrix pseudogroups},
  author={Woronowicz, Stanis{\l}aw L},
  journal={Communications in Mathematical Physics},
  volume={111},
  number={4},
  pages={613--665},
  year={1987},
  publisher={Springer}
}

@article{fuchs2019full,
  title={Full Logarithmic Conformal Field theory—an Attempt at a Status Report: LMS/EPSRC Durham Symposium on Higher Structures in M-Theory},
  author={Fuchs, J{\"u}rgen and Schweigert, Christoph},
  journal={Fortschritte der Physik},
  volume={67},
  number={8-9},
  pages={1910018},
  year={2019},
  publisher={Wiley Online Library}
}

@article{marien2017condensation,
  title={Condensation-driven phase transitions in perturbed string nets},
  author={Mari{\"e}n, Micha{\"e}l and Haegeman, Jutho and Fendley, Paul and Verstraete, Frank},
  journal={Physical Review B},
  volume={96},
  number={15},
  pages={155127},
  year={2017},
  publisher={APS}
}

@article{vanhove2022topological,
  title         = {Topological aspects of the critical three-state Potts model},
  author        = {Vanhove, Robijn and Lootens, Laurens and Tu, Hong-Hao
                   and Verstraete, Frank},
  journal       = {Journal of Physics A: Mathematical and Theoretical},
  volume        = {55},
  pages         = {235002},
  year          = {2022},
  doi           = {10.1088/1751-8121/ac68b1},
  eprint        = {2107.11177},
  archivePrefix = {arXiv},
}

@article{aasen2020topological,
  title         = {Topological Defects on the Lattice: Dualities and
                   Degeneracies},
  author        = {Aasen, David and Fendley, Paul and Mong, Roger S. K.},
  year          = {2020},
  eprint        = {2008.08598},
  archivePrefix = {arXiv},
  primaryClass  = {cond-mat.stat-mech},
}

@article{behrend1998integrable,
  title={Integrable boundaries, conformal boundary conditions and ADE fusion rules},
  author={Behrend, Roger E and Pearce, Paul A and Zuber, Jean-Bernard},
  journal={Journal of Physics A: Mathematical and General},
  volume={31},
  number={50},
  pages={L763},
  year={1998},
  publisher={IOP Publishing}
}

@article{fendley2021integrability,
  title={Integrability and braided tensor categories},
  author={Fendley, Paul},
  journal={Journal of Statistical Physics},
  volume={182},
  number={2},
  pages={1--25},
  year={2021},
  publisher={Springer}
}

@misc{liu2025tradingmathematicalphysicalsimplicity,
      title={Trading Mathematical for Physical Simplicity: Bialgebraic Structures in Matrix Product Operator Symmetries}, 
      author={Yuhan Liu and Andras Molnar and Xiao-Qi Sun and Frank Verstraete and Kohtaro Kato and Laurens Lootens},
      year={2025},
      eprint={2509.03600},
      archivePrefix={arXiv},
      primaryClass={quant-ph},
      url={https://arxiv.org/abs/2509.03600}, 
}

@article{gabai2025quantum,
  author        = {Gabai, Barak and Gorbenko, Victor and Qiao, Jiaxin
                   and Zan, Bernardo and Zhabin, Aleksandr},
  title         = {Quantum Groups as Global Symmetries},
  journal       = {J. High Energy Phys.},
  volume        = {08},
  pages         = {087},
  year          = {2025},
  doi           = {10.1007/JHEP08(2025)087},
  eprint        = {2410.24142},
  archivePrefix = {arXiv},
  primaryClass  = {hep-th},
}

@article{gabai2025quantumII,
  author        = {Gabai, Barak and Gorbenko, Victor and Qiao, Jiaxin
                   and Zan, Bernardo and Zhabin, Aleksandr},
  title         = {Quantum Groups as Global Symmetries II. Coulomb Gas
                   Construction},
  year          = {2024},
  eprint        = {2410.24143},
  archivePrefix = {arXiv},
  primaryClass  = {hep-th},
}

@book{kerler2001nonsemisimple,
  author        = {Kerler, Thomas and Lyubashenko, Volodymyr V.},
  title         = {Non-Semisimple Topological Quantum Field Theories for
                   3-Manifolds with Corners},
  series        = {Lecture Notes in Mathematics},
  volume        = {1765},
  publisher     = {Springer},
  address       = {Berlin, Heidelberg},
  year          = {2001},
  doi           = {10.1007/b82618},
}

@article{derenzi2022threedimensional,
  author        = {De Renzi, Marco and Gainutdinov, Azat M. and Geer, Nathan
                   and Patureau-Mirand, Bertrand and Runkel, Ingo},
  title         = {3-Dimensional {TQFTs} from non-semisimple modular
                   categories},
  journal       = {Selecta Math. New Ser.},
  volume        = {28},
  pages         = {42},
  year          = {2022},
  doi           = {10.1007/s00029-021-00737-z},
  eprint        = {1912.02063},
  archivePrefix = {arXiv},
}

@article{gainutdinov2018modularization,
  author        = {Gainutdinov, Azat M. and Lentner, Simon and Ohrmann, Tobias},
  title         = {Modularization of small quantum groups},
  year          = {2018},
  eprint        = {1809.02116},
  archivePrefix = {arXiv},
  primaryClass  = {math.QA},
}

\clearpage
\onecolumngrid
\appendix
\section{Appendix A: Commutant of the representation and the resulting central projectors}
\label{app:A}
The vertical product of two MPO tensors yields a new MPO tensor with bond dimension $d^2$. For each pair of fixed physical indices there is a  $d^2\times d^2$ matrix:  
\begin{align}
    A &=  M^{00} = a_1 \otimes a_2 + b_1 \otimes c_2 \\
    B &=  M^{01} = a_1 \otimes b_2 + b_1 \otimes d_2 \\
    C &=  M^{10} = c_1 \otimes a_2 + d_1 \otimes c_2 \\
    D &=  M^{11} = c_1 \otimes b_2 + d_1 \otimes d_2 
\end{align}
For the sake of the argument, let us restrict ourselves to the case $d=3$, although it can readily be generalized to any other odd number. The coproduct preserves the fundamental commutation relations defining the bialgebra \eqref{rttrelation}, so one can compute that:
\begin{align}
    [A, B^2C] &= [B, B^2C] = [C, B^2C] = [D, B^2C] = 0 \\
    [A, BC^2] &= [B, BC^2] = [C, BC^2] = [D, BC^2] = 0
\end{align}
$Q_1=B^2C$ and $Q_2=BC^2$ belong to the commutant $Z(\mathcal{A})$ of the algebra. Also, we will later see that the representation $A, B, C, D$ decomposes into a direct sum of $d$ representations of the original $a, b, c, d$ form (see Eq. \eqref{superposition}). The original representations are irreducible, and inequivalent for distinct $k$. Therefore, the dimension of the commutant is $d$. 

Furthermore, it holds that $B^3=f_1(x_1, x_2, y_1, y_2, z_1, z_2) \mathds{1}$ and $C^3=f_2(x_1, x_2, y_1, y_2, z_1, z_2)\mathds{1}$, i.e. they are both proportional to the identity (a complication will arise when $B$ or $C$ is nilpotent; we will come back to this case later). We will suppress the dependence of $f_1, f_2$ on the matrix elements and write simply $f_1, f_2$. It follows that $Q_1^2=f_1 Q_2$, $Q_2^2=f_2  Q_1$ and $Q_1Q_2=Q_2Q_1=f_1f_2\mathds{1}$. We additionally assume that $\mathds{1}, Q_1, Q_2$ are linearly independent -- we will come back to the linearly dependent case later. 

Using those algebraic relations, one can solve the following equation for the central projectors $P_i = \alpha \mathds{1} + \beta Q_1 + \gamma Q_2$ such that $P_i^2 = P_i$. We require the projectors to be non-trivial, i.e. not $0$ or $\mathds{1}$. 
\begin{align}
\label{appA:projectorCoeff1}
    &(\alpha^2 + 2f_1f_2 \beta \gamma - \alpha) \mathds{1} \; + \\
\label{appA:projectorCoeff2}
    &(\gamma^2 f_2 + 2 \alpha \beta - \beta) B^2C \; + \\
\label{appA:projectorCoeff3}
    &(\beta^2 f_1 + 2 \alpha \gamma - \gamma) BC^2 = 0
\end{align}
Linear independence requires all of the coefficients to be 0. We hence get a system of quadratic equations for $\alpha, \beta, \gamma$. Multiplying the first of these equations by $\beta \gamma$ and substituting $f_1$ and $f_2$ obtained from the second and the third, we get:
\begin{equation}
    \alpha^2 \beta \gamma + 2(1-2\alpha)^2 \beta \gamma - \alpha \beta \gamma = 0.
\end{equation}
This is trivially met if $\beta \gamma = 0$. Firstly, let us focus on the case where $\beta \gamma \neq 0$. Then, $\alpha = \frac{1}{3}$ or $\alpha = \frac{2}{3}$. Solving further for the projectors gives:
\begin{align}
    P_i^{(1)} &= \frac{1}{3}\left(\mathds{1}+\frac{q^i}{f_1^{2/3}f_2^{1/3}}Q_1+\frac{q^{2i}}{f_1^{1/3}f_2^{2/3}}Q_2\right) \\
    P_i^{(2)} &= \frac{1}{3}\left(2\mathds{1}-\frac{q^i}{f_1^{2/3}f_2^{1/3}}Q_1-\frac{q^{2i}}{f_1^{1/3}f_2^{2/3}}Q_2\right) = \mathds{1} - P_i^{(1)}. 
\end{align}
We thus get a family of 3 linearly independent projectors, each corresponding to a different choice of the cube root of unity $q^i$. One can directly check that $P_i^{(1)} P_j^{(1)} = \delta_{ij} P_i^{(1)}$, so they are orthogonal. Hence, their images are the invariant subspaces of the representation. A change of basis matrix can be constructed by choosing a basis of $\text{im} \; P_i^{(1)}$ for each of them (e.g. using SVD) and inverting the resultant matrix. 

From now, denote $P_i := P_i^{(1)}$. It remains to show that the projectors are not gauge-equivalent, i.e. for any pair of $i, j \in \{0, ..., d-1\}$ such that $i \neq j$ there exists no $X$ such that:
\bea
X P_i A P_i X^{-1} &= P_j A P_j \\
X P_i B P_i X^{-1} &= P_j B P_j \\
X P_i C P_i X^{-1} &= P_j C P_j \\
X P_i D P_i X^{-1} &= P_j D P_j.
\eea
Note that the generators $A, B, C, D,$ and the projectors $P_i, P_j$ all depend on $x_1, y_1, z_1, x_2, y_2, z_2$, so $X = X(x_1, y_1, z_1, x_2, y_2, z_2)$. We will suppress the dependence throughout.

Using the following notation for the projected generators:
\begin{eqnarray}
    P_i A P_i = A_i, \quad P_i B P_i = B_i, \quad 
    P_i C P_i = C_i, \quad P_i D P_i = D_i, \quad 
\end{eqnarray}
we require simply: 
\begin{equation}
X A_i X^{-1} = A_j, \quad X B_i X^{-1} = B_j, \quad
X C_i X^{-1} = C_j, \quad
X D_i X^{-1} = D_j.
\end{equation}
For any $x_1, y_1, z_1, x_2, y_2, z_2$, the generators satisfy the quantum determinant relation:
\begin{eqnarray}
    AD - q^{-1} BC = \mathds{1}.
\end{eqnarray}
Acting with $P_i$ on both sides yields:
\begin{equation}
    \begin{aligned}
        P_i (AD - q^{-1} BC) P_i &= P_i^2 \\
        P_i^2 (AD - q^{-1} BC) P_i^2 &= P_i \\
        (P_i A P_i) (P_i D P_i) - q^{-1} (P_i B P_i) (P_i C P_i) &= P_i \\
        A_i D_i - q^{-1} B_i C_i &= P_i,
    \end{aligned}
    \label{eq:ProjectedQuantumDeterminant}
\end{equation}
where the third equality follows from centrality of $P_i$. Similarly, we have that $A_j D_j - q^{-1} B_j C_j = P_j$. Conjugating \eqref{eq:ProjectedQuantumDeterminant} with $X$ gives:
\begin{equation}
    \begin{aligned}
        X P_i X^{-1}  &= X (A_i D_i - q^{-1} B_i C_i) X^{-1} \\
        &= (X A_i X^{-1})(X D_i X^{-1}) - q^{-1} (X B_i X^{-1}) (X C_i X^{-1}) \\
        &= A_j D_j - q^{-1} B_j C_j \\
        &= P_j
    \end{aligned}
    \label{eq:GaugeInequivalenceOfProjectorsCondition1}
\end{equation}

Alternatively, we can study the action of conjugation by $X$ on $R$, where:
\bea
R = \frac{Q_1}{f_1^{2/3} f_2^{1/3}} = \frac{B^2 C}{f_1^{2/3} f_2^{1/3}},
\eea
such that
\begin{eqnarray}
    P_k &= \frac{1}{3} (\mathds{1} + q^k R + q^{2k}R^2).
\end{eqnarray}
Projection of $R$ by $P_i$ yields:
\begin{equation}
    \begin{aligned}
R P_k &= \frac{1}{3} (R + q^k R^2 + q^{2k} \mathds{1}) \\
&= q^{2k} P_k \\
&= q^{-k} P_k \\
&\implies P_k R P_k = q^{-k} P_k.
    \end{aligned}
\end{equation}
Projecting $R$ onto the $i^{\text{th}}$ subspace and conjugating with $X$ gives:
\begin{equation}
\begin{aligned}
    X P_i R P_i X^{-1} &= \frac{1}{f_1^{2/3} f_2^{1/3}} X P_i B^2 C P_i X^{-1} \\
    &= \frac{1}{f_1^{2/3} f_2^{1/3}} (X P_i B P_i X^{-1}) (X P_i B P_i X^{-1}) (X P_i C P_i X^{-1}) \\
    &= \frac{1}{f_1^{2/3} f_2^{1/3}} (P_j B P_j) (P_j B P_j) (P_j C P_j) \\
    &= \frac{1}{f_1^{2/3} f_2^{1/3}} P_j B^2C P_j \\
    &= P_j R P_j \\
    &= q^{-j} P_j,
\end{aligned}
\end{equation}
where again we used the centrality of $P_i$ to obtain the second equality. On the other hand, we have that $X P_i R P_i X^{-1} = X (q^{-i}P_i)X^{-1}$. Combining the two gives:
\begin{equation}
    X P_i X^{-1} = q^{i-j} P_j.
    \label{eq:GaugeInequivalenceOfProjectorsCondition2}
\end{equation}
Combining \eqref{eq:GaugeInequivalenceOfProjectorsCondition1} with \eqref{eq:GaugeInequivalenceOfProjectorsCondition2} we get:
\begin{equation}
    q^{i-j} = 1,
\end{equation}
which is false when $q$ is a primitive root of unity unless $i = j$. Hence, the projectors $P_i$ are not gauge equivalent and produce gauge-inequivalent invariant subspaces.

Now, let us return to the special cases. If both $\beta$ and $\gamma$ are $0$, then the only remaining projectors are $0$ and $\mathds{1}$, so we will disregard it. 
If $\beta = 0, \gamma \neq 0$, direct substitution into \eqref{appA:projectorCoeff1} - \eqref{appA:projectorCoeff3} gives:
\begin{equation}
\begin{aligned}
    \alpha^2 - \alpha &= 0 \implies \alpha \in \{0, 1\},\\
    \gamma^2 f_2 &= 0, \\
    (2 \alpha - 1) \gamma &= 0 \implies \alpha = \frac{1}{2},
\end{aligned}
\end{equation}
which leads to a contradiction. Similarly, if $\beta \neq 0, \gamma = 0$, direct substitution into \eqref{appA:projectorCoeff1} - \eqref{appA:projectorCoeff3} gives:
\begin{equation}
\begin{aligned}
    \alpha^2 - \alpha &= 0 \implies \alpha \in \{0, 1\},\\
    (2 \alpha - 1) \beta &= 0 \implies \alpha = \frac{1}{2}, \\
    \beta^2 f_1 &= 0,
\end{aligned}
\end{equation}
which again leads to a contradiction. Hence, these two cases are impossible if $\mathds{1}, Q_1, Q_2$ are linearly independent.

The last remaining assumption to resolve is the linear independence of $\{ \mathds{1}, B^2C, BC^2\}$. Suppose they are not linearly independent, i.e.
\begin{eqnarray}
\label{appA:linearDependece}
    a \mathds{1} + b B^2C + c BC^2 = 0
\end{eqnarray}
for some constants $a,b,c \in \mathds{C}$. Direct calculation of the traces $\Tr(B^2C), \Tr(BC^2)$ gives $\Tr(B^2C) = \Tr(BC^2) = 0$. Hence, taking the trace of \eqref{appA:linearDependece}, we immediately get $a = 0$. Hence, $b B^2C + c BC^2 = 0$. Three cases arise. If $b \neq 0, c = 0$:
\begin{equation}
    B^2C = 0 \implies B^3C = f_1 C = 0 \implies C = 0 \lor f_1 = 0, 
\end{equation}
i.e. either $C=0$ or $B$ is nilpotent.
Analogously, $b = 0, c \neq 0$ leads to:
\begin{equation}
    BC^2 = 0 \implies BC^3 = f_2 B = 0 \implies B = 0 \lor f_2 = 0, 
\end{equation}
i.e. either $B=0$ or $C$ is nilpotent. If $b \neq 0, c \neq 0$ we can write $B^2C = s BC^2$ for some non-zero $s$. Multiplying by $B$ and $C$ separately gives:
\begin{equation}
\begin{aligned}
    f_1 C &= s B^2C^2 \\
    s f_2 B &= B^2 C^2 \implies f_1 C = s^2 f_2 B.
\end{aligned}
\end{equation}
Hence, either $B$ and $C$ are proportional or $f_1 = f_2 = 0$, i.e. $B$ and $C$ are both nilpotent. We will return to those cases later (see Appendix E). 

Consequently, let us investigate when proportionality is true. A trivial example of proportional $B$, $C$ is when exactly one of them is $0$ - this will be investigated further in Appendix E. The remaining case is $B = \kappa C$, where we assume the proportionality constant $\kappa \neq 0$. From the diagonal blocks of $B$ and $C$ we get:
\begin{align}
    x_1 y_2 &= \kappa \frac{z_2}{x_1} \\
    q x_1 y_2 &= \kappa \frac{z_2}{q x_1} \\
    q^2 x_1 y_2 &= \kappa \frac{z_2}{q^2 x_1} \\
    \implies y_2 &= z_2 = 0.
\end{align}
Analogously, from the $B_{1, 2}, C_{1, 2}$ block we get:
\begin{align}
    \frac{y_1}{x_2} &= \kappa x_2 z_1 \\
    \frac{y_1}{q x_2} &= \kappa q x_2 z_1 \\
    \frac{y_1}{q^2 x_2} &= \kappa q^2 x_2 z_1 \\
    \implies y_1 &= z_1 = 0.
\end{align}
Hence, $B = C = 0$, so this is also a special case with both $B$ and $C$ nilpotent. This case amounts to multiplication of MPOs constructed from two diagonal tensors, i.e.
\begin{align}
    A &= x_1 x_2 I_q \otimes I_q \\
    D &= \frac{1}{x_1 x_2} I_{q^{-1}} \otimes I_{q^{-1}}.
\end{align}
One finds that the matrix
\begin{equation}
    F = \bigoplus_{k=0}^2 \sigma^k 
\end{equation}
block-diagonalises $A, D$ into the following form:
\begin{align}
    F^{-1} A F &= \bigoplus_{k=0}^2 x_1 x_2 I_q \\
    F^{-1} D F &= \bigoplus_{k=0}^2 \frac{1}{x_1 x_2} I_{q^{-1}}. 
\end{align}
\section{Appendix B: Generic case fusion rules through diagonalising A}
In this section we explicitly calculate the change of basis matrix $U$ (built from the $\alpha$ and $\beta$ coefficients given explicitly below) which simultaneously block-diagonalises $A, B, C$ and $D$.  In this section, all indices, i.e. $i, j$ and $k$ are taken modulo $d$.

First, let us find a basis diagonalising $A$. The action of $A$ on the original $d^2$-dimensional tensor product virtual space is:
\bea
A\ket{i,j}=x_1x_2q^{i+j}\ket{i,j}+y_1z_2\ket{i-1,j-1},
\eea

Notice that the $d$ subspaces $V_k$, $k\in\{0,\ldots,d-1\}$, defined as
\bea
V_k=\text{span} \left\{\ket{i,i+k} \; | \; i \in \{0,...,d-1\} \right\},
\label{eq:appBInvariantSubspaces}
\eea
are invariant subspaces under the action of $A$. Hence, we can proceed with the further diagonalisation of $A$ within each subspace $V_k$ separately, using the basis \eqref{eq:appBInvariantSubspaces}.

Let $v_k^{\lambda} \in V_k$ be a generic linear combination of the basis vectors from \eqref{eq:appBInvariantSubspaces}:
\bea
v^{\lambda}_k=\sum_{i=0}^{d-1}\alpha^{\lambda}_{i,k}\ket{i,i+k}
\eea
For diagonalisation of A, we require $Av_k^{\lambda}=\lambda v_k^{\lambda}$. The action of $A$ on $v_k^{\lambda}$ yields:
\bea
A v^{\lambda}_k=\sum_{i=0}^{d-1
}\left(x_1x_2q^{2i+k}\alpha_{i,k}^{\lambda}+y_1z_2\alpha_{i+1,k}^{\lambda}\right)\ket{i,i+k}
\eea
Substituting into the eigenvalue equation and comparing coefficients of each basis vector gives:
\bea
x_1x_2q^{2i+k}\alpha_{i,k}^{\lambda}+y_1z_2\alpha_{i+1,k}^{\lambda} = \lambda \alpha_{i,k}^{\lambda},
\eea
which leads to the recursion relation on the $\alpha$ coefficients:
\bea
\frac{\alpha_{i+1,k}^{\lambda}}{\alpha_{i,k}^{\lambda}}=\frac{\lambda-x_1x_2q^{2i+k}}{y_1z_2}.
\label{eq:appBRecursionOnAlpha}
\eea
As all the indices are taken modulo $d$, using the recursion relation $d$ times leads to an equation for $\lambda$ in terms of the original variables $x_1, x_2, y_1, z_2$:
\bea
(y_1z_2)^d=\prod_{i=1}^{d}\left({\lambda}-x_1x_2q^{2i+k}\right)\label{eq:appBLambda}.
\eea
Expanding the product on the right-hand side we get:
\bea
(y_1z_2)^d&=&{\lambda}^d+{\lambda}^{d-1}(-x_1x_2q^k)\sum_iq^{2i}+\lambda^{d-2}(-x_1x_2q^k)^2\sum_{i_1<i_2}q^{2i_1+2i_2}\\&&+\ldots+(-x_1x_2q^k)^d\sum_{i_1<...< i_d}q^{2i_1+...+2i_d}.
\eea
Using the fact that, for odd $d$, $q^2$ is also a primitive
$d^{\mathrm{th}}$ root of unity, we have
\begin{equation}
    \prod_{i=0}^{d-1} (x-q^{2i}) = x^d-1.
\end{equation}
Expanding the left hand side and comparing coefficients leads to the identity
\bea
\sum_{0 \leq i_1<\ldots<i_m \leq d-1}q^{2(i_1+\ldots+i_m)}=
\left\{ \begin{array}{rcl}
         0 & \text{if}
         & 1 \leq m \leq d-1  \\
         1& \mbox{if} & m=0 \lor m=d
                \end{array}\right.
\eea
for $d$ odd. From this, only the first term and the last term in the expansion are non-zero. We get simply
\bea
{\lambda}^d=(x_1x_2)^d+(y_1z_2)^d.\label{eq:appBLambdaSimplified}
\eea
Importantly, the eigenvalues are independent of $k$. This means that each $\lambda$ is $d$-fold degenerate, and appears exactly once in each $V_k$. Performing additional changes of basis on any of the eigenspaces of $A$, $E_\lambda^A$, will leave $A$ invariant - we will use this to simultaneously diagonalise $A$ and block-diagonalise $B$ and $C$.

The coefficients $\alpha_{i,k}^{\lambda}$ can be calculated directly from Eq. \eqref{eq:appBRecursionOnAlpha}, given $\lambda$ and a choice of normalisation by fixing one of the $\alpha_{i,k}^{\lambda}$. If we take as a convention $\alpha_{0,k}^{\lambda}=1$ then 
\bea
\alpha_{i,k}^\lambda=\frac{1}{(y_1z_2)^{i}}\prod_{p=0}^{i-1}\left({\lambda}-x_1x_2q^{2p+k}\right).
\label{eq:AlphaExplicitForm}
\eea
\subsection{Simultaneous block-diagonalisation of $B$ and $C$ on $A$-eigenspaces}
From the original RTT relations in Eq. \eqref{rttrelation} and the fact that the coproduct respects those relations, we get
\begin{eqnarray}
    ABv &= q^{-1}BAv, \\
    ACv &= q^{-1}CAv,  
\end{eqnarray}
from which it follows that $B$ and $C$ cycle between eigenspaces of $A$:
\bea
B E_\lambda^A &\subseteq E_{q^{-1}\lambda}^A, \\
C E_\lambda^A &\subseteq E_{q^{-1}\lambda}^A.
\eea
We will hence search for a basis where $B$ and $C$ are a direct sum of $d$ weighted permutation matrices.

In fact, on each cycle $B$ and $C$ must be proportional to each other. To see that, let $\{u_{0}, ..., u_{d-1}\}$ be a cycle such that $Bu_{i} = \mu_{i}^B u_{i+1}$, and $Cu_{i} = \mu_{i}^C u_{i+1}$, where $\mu_{i}^B, \mu_{i}^C \in \mathds{C}$ are the weights of the cyclic shift. Then:
\bea
BC u_i &= \mu_{i+1}^B \mu_{i}^C u_{i+2}, \\
CB u_i &= \mu_{i+1}^C \mu_{i}^B u_{i+2}.
\eea
As $[B, C] = 0$, it follows that the proportionality ratio between $B$- and $C$-weights is equal for all $i$:
\bea
\frac{\mu_{i}^C}{\mu_{i}^B} = \frac{\mu_{i+1}^C}{\mu_{i+1}^B} \quad \forall_{i}
\eea
Therefore, if a basis splitting the space into a direct sum of $d$ cycles exists, a necessary (and sufficient) condition it must fulfil on each cycle is:

\bea
Av^{\lambda,\mu}=\lambda v^{\lambda,\mu},\quad Bv^{\lambda,\mu}=\mu Cv^{\lambda,\mu},\ \text{with}\ v^{\lambda,\mu}=\sum_{k=0}^{d-1}\beta^{\lambda, \mu}_k v_k^{\lambda}  \quad \text{for some} \; \{\beta^{\lambda, \mu}_k\} \label{eq:betasDefiningEquation}
\eea
where $\mu := \frac{\mu^B_i}{\mu^C_i}$ for all $i$. Note that $\mu$ can differ between the $d$ cycles, so there are $d$ possible values of $\mu$ in general. We first calculate the action of $B$ and $C$ on our basis vectors $v_k^{\lambda}$
\bea
C v_k^{\lambda}&=&\frac{z_2}{x_1}v_{k-1}^{q^{-1}{\lambda}}+\frac{z_1}{x_1y_1z_2}{\lambda}q^{-1}({\lambda}-x_1x_2q^k)v_{k+1}^{q^{-1}{\lambda}}\\
B v_k^{\lambda}&=&\frac{y_2}{x_2}{\lambda}q^{-k}v_{k-1}^{q^{-1}{\lambda}}+\frac{1}{x_2z_2}q^{-k-1}({\lambda}-x_1x_2q^k)v_{k+1}^{q^{-1}{\lambda}}.
\eea
Action on $v^{{\lambda},{\mu}}$ then gives
\bea
\label{appB:BAction}
Bv^{{\lambda},{\mu}}&=&\sum_{k=0}^{d-1}\left(\beta_{k+1}^{{\lambda},{\mu}}\frac{y_2}{x_2}{\lambda}q^{-k-1}+\beta_{k-1}^ {{\lambda},{\mu}}\frac{1}{x_2z_2}q^{-k}({\lambda}-x_1x_2q^{k-1})\right)v_k^{q^{-1}{\lambda}}\\
\label{appB:CAction}
Cv^{{\lambda},{\mu}}&=&\sum_{k=0}^{d-1}\left(\beta_{k+1}^{{\lambda},{\mu}}\frac{z_2}{x_1}+\beta_{k-1}^{{\lambda},{\mu}}\frac{z_1}{x_1y_1z_2}q^{-1}{\lambda}({\lambda}-x_1x_2q^{k-1})\right)v_k^{q^{-1}{\lambda}}
\eea
which leads, using \eqref{eq:betasDefiningEquation}, to the following relation
\bea
\beta_{k+1}^{{\lambda},{\mu}}\left(\frac{y_2}{x_2}{\lambda}q^{-k-1}-\frac{z_2}{x_1}\mu\right)=\frac{\beta_{k-1}^{{\lambda},{\mu}}}{z_2}\left({\lambda}-x_1x_2q^{k-1}\right)\left(\frac{z_1q^{-1}{\lambda}{\mu}}{x_1y_1}-\frac{q^{-k}}{x_2}\right)\label{betarecursion}
\eea
Normalising $\beta_{0}^{\lambda, \mu}$ to 1 gives:
\bea
\beta_{2n}^{\lambda, \mu} = \prod_{i=0}^{n-1}\frac{\frac{1}{z_2} \left({\lambda}-x_1x_2q^{2i}\right)\left(\frac{z_1q^{-1}{\lambda}{\mu}}{x_1y_1}-\frac{q^{-2i-1}}{x_2}\right)}{\frac{y_2}{x_2}{\lambda}q^{-2i-2}-\frac{z_2}{x_1}\mu}, \quad \beta_{2n+1}^{\lambda, \mu} = \beta_{2n+1+d}^{\lambda, \mu}
\label{eq:BetaExplicitForm}
\eea
for odd $d$.
Taking a product over the index $k$ leads to the following consistency equation for $\mu$:
\bea
\mu^d=\frac{x_1^d}{x_2^d}\times\frac{y_1^d+y_2^d(x_1^dx_2^d+y_1^dz_2^d)}{z_2^d+z_1^d(x_1^dx_2^d+y_1^dz_2^d)},\label{deltaeq}
\eea
which can be used to find all $d$ solutions for $\mu$.
Acting with $B$ and $C$ on $v^{{\lambda},{\mu}}$ (or directly raising $B$ and $C$ to the $d^{\text{th}}$ power, as $B^d$ and $C^d$ are proportional to the identity), one can show that 
\bea
B^d v^{{\lambda},{\mu}}={\mu_B}^dv^{{\lambda},{\mu}},\quad\text{with}\quad {\mu_B}^d=\frac{y_2^d}{x_2^d}(x_1^dx_2^d+y_1^dz_2^d)+\frac{y_1^d}{x_2^d}\label{yeq}
\eea
and
\bea
C^d v^{{\lambda},{\mu}}={\mu_C}^dv^{{\lambda},{\mu}},\quad\text{with}\quad {\mu_C}^d=\frac{z_1^d}{x_1^d}(x_1^dx_2^d+y_1^dz_2^d)+\frac{z_2^d}{x_1^d}\label{zeq}
\eea
which can be used as a consistency check for $\mu$, i.e. $\mu^d=\frac{\mu_B^d}{\mu_C^d}$.

\subsection{Fusion rules}
In the end, we find that there are $d$ blocks, each parametrised by a root of $\mu^d$ obtained from \eqref{deltaeq}. Equivalently, the blocks can be parametrised by a root of  $\mu_B^d$ or $\mu_C^d$. We thus obtain $d$ distinct blocks, in agreement with the similarity transformations mentioned in the main text. Using equations \eqref{eq:appBLambdaSimplified}, \eqref{yeq} and \eqref{zeq} we observe that
\bea
\left( \begin{array}{cc} {\lambda}^d & {\mu_B}^d \\
{\mu_C}^d & \tau^d \end{array} \right)=
\left( \begin{array}{cc} x_1^d & y_1^d \\
z_1^d & t_1^d \end{array} \right).
\left( \begin{array}{cc} x_2^d & y_2^d \\
z_2^d & t_2^d \end{array} \right)\eea
with $\tau$, $t_1$, $t_2$ chosen such that the determinant of each matrix is one. At this point, we need to choose ${\lambda}$, ${\mu_B}$ and ${\mu_C}$ on a consistent branch of the $d$-th root, as mentioned in the main text. Defining $g_{12}=(\td{x}_{12},\td{y}_{12},\td{z}_{12})=({\lambda}^d,{\mu_B}^d,{\mu_C}^d)$ leads to equation \eqref{su2rule} in the main text.
\subsection{${}^1\!F$ definition}
To summarise, using the notation of the main text, we have
\bea
\left({^1F}^{(g_1, k_1) (g_2, k_2)}_{(g_1g_2, k_{12})}\right)_{l, {(\lambda, \mu)}} = \alpha_{i,k}^{\lambda} \beta_{k}^{\lambda, \mu},\qquad l=di+j,\quad k=j-i \bmod d,
\eea
where the original $d^2$-dimensional basis $e_l$ is labelled by $l$ s.t. $\ket{i, j} = e_{di+j}$, and the outgoing basis is labelled by the eigenspace of $A$, i.e. $\lambda$, and the solution to \eqref{eq:betasDefiningEquation}, i.e. $\mu$. The $\alpha$ and $\beta$ coefficients are given by \eqref{eq:AlphaExplicitForm} and \eqref{eq:BetaExplicitForm} respectively. 

\section{Appendix C: Solving for the Jordan-block basis in the A nilpotent case for $d=3$}
Let us again focus on the case with $d=3$. If $A^3 = 0$, then $(x_1 x_2)^3 + (y_1 z_2)^3 = 0$, as $A^3 = ((x_1 x_2)^3 + (y_1 z_2)^3) \mathds{1}$. Up to a gauge transformation, the solution is $x_1 x_2 = -y_1 z_2$. $A$ becomes non-diagonalisable, and acquires a non-zero kernel. As $A$ is non-diagonalisable, the analysis in Appendix B is not valid. One can however search for Jordan  chains, which turn out to come in multiplicity of 3. Indeed, $A$ has rank 6, $A^2$ has rank 3, and $A^3$ has rank 0. 

We are thus searching for $v_i \in \ker A, u_i \in \ker A^2 \setminus \ker A, t_i \in \ker A^3\setminus \ker A^2, \; i\in \{0,1,2\}$ such that:
\begin{align}
\label{appB:A_adapted_basis_1}
    A v_i &= 0, \\
\label{appB:A_adapted_basis_2}
    A u_i &= v_i, \\
\label{appB:A_adapted_basis_3}
    A t_i &= u_i.
\end{align}
Note that $\ker A, \ker A^2, \ker A^3$ are $B$-stable and $C$-stable, i.e.:
\begin{align}
\forall v \in \ker A, ABv = q^{-1}BAv = 0 &\implies Bv \in \ker A, \\
\forall u \in \ker A^2 \; \text{s.t.} Au = v', ABu = q^{-1}BAu = q^{-1}Bv' &\implies Bu \in \ker A^2 \\
\forall t \in \ker A^3  &\implies Bt \in \ker A^3,
\end{align}
where the last implication is true simply because $\ker A^3 = V$. Identical relations hold for $C$-stability. Therefore, we can diagonalise $B$ or $C$ on the quotients $\ker A^{r+1} / \ker A^{r}$ independently (see Appendix D for the detailed proof). Whether they can be simultaneously diagonalised depends on additional compatibility conditions, which will turn out to be true in this case. 

We will proceed in the following way: first, we will find a basis meeting the conditions in \eqref{appB:A_adapted_basis_1} - \eqref{appB:A_adapted_basis_3}. Then, we will diagonalise B on $\ker A$ and the quotient spaces $\ker A^2 / \ker A$, $\ker A^3 / \ker A^2$ and solve for the counterterms to eliminate upper-diagonal contributions. 

A convenient basis meeting the criteria in \eqref{appB:A_adapted_basis_1} - \eqref{appB:A_adapted_basis_3} arises by taking 3 original basis vectors, and 3 columns of $A, A^2$. Rescaling for convenience, let us define: 
\[
\begin{aligned}
v_0 &= \begin{pmatrix}
1\\0\\0\\0\\1\\0\\0\\0\\q^2
\end{pmatrix},
&
v_1 &= \begin{pmatrix}
0\\1\\0\\0\\0\\q\\q\\0\\0
\end{pmatrix},
&
v_2 &= \begin{pmatrix}
0\\0\\1\\q^2\\0\\0\\0\\1\\0
\end{pmatrix},
\\[2em]
u_0 &= \begin{pmatrix}
0\\0\\0\\0\\-\dfrac{1}{x_1x_2}\\0\\0\\0\\\dfrac{q}{x_1x_2}
\end{pmatrix},
&
u_1 &= \begin{pmatrix}
0\\0\\0\\0\\0\\-\dfrac{1}{x_1x_2}\\\dfrac{q^2}{x_1x_2}\\0\\0
\end{pmatrix},
&
u_2 &= \begin{pmatrix}
0\\0\\0\\-\dfrac{1}{x_1x_2}\\0\\0\\0\\\dfrac{1}{x_1x_2}\\0
\end{pmatrix},
\\[2em]
t_0 &= \begin{pmatrix}
0\\0\\0\\0\\0\\0\\0\\0\\\dfrac{1}{x_1^2x_2^2}
\end{pmatrix},
&
t_1 &= \begin{pmatrix}
0\\0\\0\\0\\0\\0\\\dfrac{1}{x_1^2x_2^2}\\0\\0
\end{pmatrix},
&
t_2 &= \begin{pmatrix}
0\\0\\0\\0\\0\\0\\0\\\dfrac{1}{x_1^2x_2^2}\\0
\end{pmatrix}.
\end{aligned}
\]

Starting from $\ker A$, the action of $B, C$ on $v_i$ is:
\begin{equation}
     B v_i =\frac{q^2 y_1}{x_2} v_{i+1}, \hspace{20pt} C v_i =\frac{-x_2}{y_1} v_{i-1}.
\end{equation}
These are cyclic shifts, so $B, C$ can be simultaneously diagonalised by the Fourier basis:
\begin{equation}
    \tilde{v}_{\omega} = v_0 + \omega^{-1} v_1 + \omega^{-2} v_2, 
\end{equation}
where $\omega \in \{1, q, q^2\}$. The action of $B, C$ on $\tilde{v}_{\omega}$ is:
\begin{equation}
     B \tilde{v}_{\omega} = \lambda_{\omega}^B  \tilde{v}_{\omega}, \hspace{20pt} C \tilde{v}_{\omega} = \lambda_{\omega}^C  \tilde{v}_{\omega},
\end{equation}
with eigenvalues:
\begin{equation}
    \lambda_{\omega}^B = \frac{q^2 y_1}{x_2} \omega, \hspace{20pt} \lambda_{\omega}^C = \frac{-x_2}{y_1} \omega^{-1}.
\end{equation}
Now, compute the action of $B, C$ on $\{u_i\}$.
\begin{align}
B u_0
&= \frac{q y_1}{x_2}\,u_1
   -\frac{q^2 y_1}{x_1x_2^2}\,v_1
   +\frac{y_2}{x_2}\,v_2,
\\
B u_1
&= \frac{q y_1}{x_2}\,u_2
   +\frac{q^2 y_2}{x_2}\,v_0
   -\frac{q y_1}{x_1x_2^2}\,v_2,
\\
B u_2
&= \frac{q y_1}{x_2}\,u_0
   -\frac{y_1}{x_1x_2^2}\,v_0
   +\frac{q y_2}{x_2}\,v_1.
\end{align}

\begin{align}
C u_0
&= -\frac{q^2 x_2}{y_1}\,u_2
   +\frac{q^2 z_1}{x_1}\,v_1,
\\
C u_1
&= -\frac{q^2 x_2}{y_1}\,u_0
   +\frac{z_1}{x_1}\,v_2,
\\
C u_2
&= -\frac{q^2 x_2}{y_1}\,u_1
   +\frac{q z_1}{x_1}\,v_0.
\end{align}
Note that there is a contribution from $v_i$, so in order to block-diagonalise $B$ and $C$ we have to subtract kernel vectors from $\{u_i\}$. On the quotient space $\ker A^2 /\ \ker A$, the induced actions are again cyclic shifts. Define as before:
 
\begin{equation}
    \tilde{u}_{\omega} = u_0 + \omega^{-1} u_1 + \omega^{-2} u_2,
\end{equation}
where $\omega \in \{1, q, q^2\}$. The action of $B, C$ on $\tilde{u}_{\omega}$ is:
\begin{equation}
    B \tilde{u}_{\omega} = \mu_{\omega}^B \tilde{u}_{\omega}, \hspace{20pt}
    C \tilde{u}_{\omega} = \mu_{\omega}^C \tilde{u}_{\omega},
\end{equation}
with eigenvalues:
\begin{equation}
    \mu_{\omega}^B = \frac{q y_1}{x_2}\,\omega, \hspace{20pt}
    \mu_{\omega}^C = -\frac{q^2 x_2}{y_1}\,\omega^{-1}.
\end{equation}
Define: $\tilde{u}_i := \tilde{u}_{q^i}$, $\tilde{v}_i := \tilde{v}_{q^i}$. From the action on $\{u_i\}$ we can get the action on $\{\tilde{u}_i\}$, and reexpress it in terms of $\{\tilde{v}_i\}$.  Suppose:
\begin{align}
    B \tilde{u}_i &= \mu^B_i \tilde{u}_i + \sum_{j=0}^2 e_j \tilde{v}_j, \\
    C \tilde{u}_i &= \mu^C_i \tilde{u}_i + \sum_{j=0}^2 f_j \tilde{v}_j,
\end{align}
Then let $\hat{u}_i = \tilde{u}_i + v', \; v' \in \ker A, \; v' = \sum_{j=0}^2 g_{ij} \tilde{v}_j$, and require $B \hat{u}_i = \mu_i^B \hat{u}_i$. This implies
\begin{align}
    B \hat{u}_i &= B (\tilde{u}_i + v') \\
    &= B\tilde{u}_i + \sum_{j=0}^2 g_{ij} B \tilde{v}_j \\
    &= \mu_i^B \tilde{u}_i + \sum_{j=0}^2 e_j \tilde{v}_j + \sum_{j=0}^2 g_{ij} \lambda_{j}^B \tilde{v}_{j} \\ &\implies e_j = g_{ij} (\mu_i^B - \lambda_j^B)
\end{align}
Analogously, 
\begin{align}
    C \hat{u}_i &= C (\tilde{u}_i + v') \implies f_j = g_{ij} ( \mu_i^C - \lambda_j^C).
\end{align}
Direct check shows that the two conditions are equivalent, and thus it is possible to solve for $g_{ij}$. The solution is:
\begin{equation}
\left(g_{ij}\right)=
\begin{pmatrix}
0 &
-\dfrac{q x_1 x_2 y_2 + x_1 x_2 y_2 + y_1}{x_1 x_2 y_1 (q-1)} &
-\dfrac{y_1 z_1}{x_1 x_2 (q-1)}
\\[1.2em]
\dfrac{y_1 z_1}{x_1 x_2 (2q+1)} &
0 &
\dfrac{x_1 x_2 y_2 - y_1}{x_1 x_2 y_1 (q-1)}
\\[1.2em]
\dfrac{q x_1 x_2 y_2 - y_1}{x_1 x_2 y_1 (q-1)} &
-\dfrac{q\,y_1 z_1}{x_1 x_2 (q-1)} &
0
\end{pmatrix}.
\end{equation}
Lastly, the action of $B, C$ on $t_i$ is:
\begin{align}
B t_0
&= \frac{y_1}{x_2}\,t_1
   -\frac{q y_1}{x_1x_2^2}\,u_1
   +\frac{q^2 y_2}{x_2}\,u_2,
\\
B t_1
&= \frac{y_1}{x_2}\,t_2
   +\frac{q y_2}{x_2}\,u_0
   -\frac{y_1}{x_1x_2^2}\,u_2,
\\
B t_2
&= \frac{y_1}{x_2}\,t_0
   -\frac{q^2 y_1}{x_1x_2^2}\,u_0
   +\frac{y_2}{x_2}\,u_1.
\end{align}

One can perform a Fourier transform and analogously solve for $\hat{t}_i$ such that $B \hat{t}_i = \nu_i^B \hat{t}_i$, with eigenvalues:
\begin{equation}
    \nu_{\omega}^B = \frac{y_1}{x_2}\,\omega, \hspace{20pt}
    \nu_{\omega}^C = -\frac{q x_2}{y_1}\,\omega^{-1}.
\end{equation}
As the 3 Jordan chains $\{\tilde{v}_i, \hat{u}_i, \hat{t}_i\}$ are invariant subspaces under the action of $A, B, C$, from the relation $DA = \mathds{1}+qBC$ and commutation relations of $D$ with $B,C$ it follows that $D$ must preserve the invariant subspaces as well (see Appendix D for the general argument). In total, we get:
\begin{align}
    A &= \mathds{1} \otimes J \\
    B &= \frac{q^2y_1}{x_2} I_q \otimes I_{q^{-1}} \\
    C &= -\frac{x_2}{y_1} I_{q^{-1}} \otimes I_{q^{-1}} \\
    D &= \mathds{1} \otimes \tilde{D}, \;
\tilde{D} = 
    \begin{pmatrix}
        0 & 0 & \frac{1}{3}\Lambda_{D} \\
        1-q & 0 & 0 \\
        0 & 1-q^2 & 0 \\
    \end{pmatrix},
\end{align}
where $\Lambda_{D} = \frac{1+y_1^3 z_1^3}{x_1^3 x_2^3} - \frac{y_2^3}{y_1^3}$, i.e. the unique eigenvalue of $D^3$.
Finally, notice that the outcome is only dependent on the product of 2 $SL(2)$ elements, $g_1 g_2$. To see that, suppose:
\begin{equation}
g_1 g_2 =
\begin{pmatrix}
0 & u \\
v & w
\end{pmatrix}
\end{equation}
Then
\begin{equation}
g_2
=
g_1^{-1}(g_1g_2)
=
\begin{pmatrix}
\tilde{t}_1 & -\tilde{y}_1 \\
-\tilde{z}_1 & \tilde{x}_1
\end{pmatrix}
\begin{pmatrix}
0 & u \\
v & w
\end{pmatrix}.
\end{equation}
Hence, $\tilde{x}_2 = -\tilde{y}_1 v$, i.e. $\frac{x_2}{y_1}$ is fixed by $g_1 g_2$ and the choice of gauge, i.e. $k_1, k_2$. This proves that the fused representation respects the product in $SL(2)$.
\section{Appendix D: Proof of general form in the A nilpotent case}

\subsection{Dimension of kernel in invariant subspaces of $A$: $
\dim (\ker A \cap V_k) = 1$}
As in Appendix B, let $V_k = \text{span} \{\ket{i, i+k}\}, i \in \{0, ..., d-1\}$.
Then 
\begin{equation}
\begin{aligned}
    A \ket{i, i+k} &= x_1 x_2 q^{2i+k} \ket{i, i+k} + y_1 z_2 \ket{i-1, i-1+k} \in V_k \\
    &\implies A V_k \subseteq V_k.
\end{aligned}
\end{equation}
Let $v_k = \sum_{i=0}^{d-1} \gamma_{i,k} \ket{i, i+k}$. To calculate the kernel, require $Av_k = 0$. This leads to:
\begin{equation}
\begin{aligned}
    A v_k = \sum_{i=0}^{d-1} \gamma_{i,k} A \ket{i, i+k} &= 0 \\
 \gamma_{i,k} x_1 x_2 q^{2i+k} + \gamma_{i+1,k}y_1 z_2 &= 0 \quad \forall i \\
 \frac{\gamma_{i+1,k}}{\gamma_{i,k}} &= -\frac{x_1 x_2 q^{2i+k}}{y_1 z_2}  \quad \forall i.
\end{aligned}
\end{equation}
Fixing $\gamma_{0,k}$ determines all $\gamma_{i,k}$ by proportionality constants, so for any choice of $\gamma_{0,k}$ the resulting vectors will be proportional to each other. Hence $\dim (\ker A \cap V_k) \leq 1$. 

On the other hand, $V_k$ is $A$-invariant and $A^d=0$, so the restriction $(A|_{V_k})^d=0$. Hence $\dim (\ker A \cap V_k) \geq 1$. It follows that  $\dim (\ker A \cap V_k) = 1$.

\subsection{Jordan block decomposition of A}
From Jordan's canonical form theorem, any nilpotent $d \times d$ matrix $M$ is similar to a direct sum of nilpotent Jordan blocks, $\bigoplus_{i=0}^s J_{m_i}(0)$, where $\sum_{i=0}^s m_i = d$ and $[J_m(0)]_{kl} = \delta_{k,l-1} \; \text{for} \; k \in \{0,..., m-2\}$. Each Jordan block contributes one vector to the kernel, i.e. $\dim \ker J_{m_i}(0) = 1 \quad \forall i$. It follows from $\dim (\ker A|_{V_k}) =1$ that $A|_{V_k}$ is similar to a single Jordan block, i.e. $A|_{V_k} \simeq J_{m_0}(0)$. Because $(A|_{V_k})^d=0$, $m_0 \leq d$. As $A|_{V_k}$ is a $d \times d$ matrix, it must be that $m_0 = d$. Thus,  for all $V_k$ there exists a similarity transformation $S_k$ such that $S_k^{-1} A|_{V_k} S_k = J_{d}(0)$.

\subsection{Stability of $A$-filtration under $B, C$}

From the commutation relations we have:
\begin{equation}
    A^m B = q^{-m} B A^m, \quad A^m C = q^{-m} C A^m.
\end{equation}
Suppose $v \in \ker A^r$. Then:
\begin{equation}
    A^rBv=q^{-r}BA^rv=0, \quad A^rCv=q^{-r}CA^rv=0,
\end{equation}
i.e. $Bv \in \ker A^r, \; Cv \in \ker A^r$. Therefore:
\begin{equation}
    B (\ker A^r) \subseteq (\ker A^r), \quad C (\ker A^r) \subseteq (\ker A^r).
\end{equation}
\subsection{Dimension of quotient spaces: $\dim (\ker A^{r+1} / \ker A^{r}) = d$}
From the Jordan block decomposition of A, we know that:
\begin{equation}
\begin{aligned}
    \dim \ker A^{r}
    &= \dim \ker \bigoplus_{k=0}^{d-1} (A|_{V_k})^{r} \\
    &= \sum_{k=0}^{d-1} \dim (\ker (A|_{V_k})^r) \\
    &= \sum_{k=0}^{d-1} \dim \ker J_{d}(0)^r \\
    &= \sum_{k=0}^{d-1} r = dr,
\end{aligned}
\end{equation}
where the fourth equality follows from $\dim \ker J_{m}(0)^r = r$ for $0 \leq r \leq m$. Therefore, the dimension of each quotient is equal to $d$:
\begin{equation}
    \dim  (\ker A^{r+1} / \ker A^{r}) = d(r+1) - dr = d \quad \forall r \in \{0, ..., d-1\}.
\end{equation}
\subsection{Simultaneous diagonalisation of $B, C$ on $\ker A$}
From previous analysis (see Appendix B and Appendix C) we know that $B V_{k} \subseteq V_{k+1}$ and $C V_{k} \subseteq V_{k-1}$. Choose non-zero $u_k \in \ker A \cap V_k$ and rescale them so that:
\begin{equation}
    B u_k = \beta u_{k+1}, \quad C u_k = \gamma_k u_{k-1}.
\end{equation}
This is always possible if $B, C$ are not nilpotent: since $\ker A \cap V_k$ are 1-dimensional, one can always choose a non-zero member of each space and rescale them so that the action of $B$ is just multiplication by the same constant across $k$. Since $DA = \mathds{1} + qBC$, we have:
\begin{equation}
\begin{aligned}
    DA|_{\ker A} &= \mathds{1}|_{\ker A} + qBC|_{\ker A} \\
    0 &= \mathds{1}_d + qBC|_{\ker A} \\
    BC|_{\ker A} &= -q^{-1} \mathds{1}_d \\
    \beta \gamma_k &= -q^{-1} \; \forall k.
\end{aligned}
\end{equation}
Therefore, $\gamma_k := \gamma$ is independent of $k$.
Going to the Fourier basis, denoted here $f_{l,0}$, diagonalises both cyclic matrices. Define:
\begin{equation}
    f_{l, 0} = \sum_{k=0}^{d-1} q^{-lk} u_k.
\end{equation}
It is easy to check that the action of $B,C$ becomes:
\begin{equation}
    B f_{l,0} = \beta q^{l} f_{l,0}, \quad C f_{l,0} = \gamma q^{-l} f_{l,0}. 
\end{equation}
\subsection{Induced maps on quotients and isomorphisms $\phi_r$ from $\ker A^{r+1} / \ker A^r$ to $\ker A$}
Define canonical quotient maps $\pi_r : $
\begin{equation}
    \pi_r : \ker A^{r+1} \to Q_r := \ker A^{r+1}/\ker A^r,
    \qquad
    \pi_r(v) = [v]_r := v + \ker A^r .
\end{equation}
The induced action of $\bar{B_r}, \bar{C_r}$ on the quotient spaces is
\begin{equation}
    \bar{B_r}[v]_r := [Bv]_r, \quad \bar{C_r}[v]_r := [Cv]_r.
\end{equation}
These are invertible and commute with each other. Also, define maps $\phi_r$ from each quotient $Q_r$ to $\ker A$:
\begin{equation}
    \phi_r : Q_r \to \ker A,
    \qquad
    \phi_r([v]_r) = A^{r} v.
\end{equation}
This map is well-defined despite the choice of representative $v$ being arbitrary. Indeed, suppose we choose 2 different representatives such that $[v]_r = [v']_r$. Then, $v - v' \in \ker A^r \implies A^r v - A^r v' = 0 \Leftrightarrow \phi_r([v]_r) = \phi_r([v']_r)$. It is also injective, as $\ker \phi_r = \{[v]_r  \;|\; v \in \ker A^r \} = [0]_r$ by definition of the quotient. Lastly, it is surjective, as $\dim \ker A = d$ and $\dim Q_r = d$. Hence, all $\phi_r$ are isomorphisms and one can define their inverses $\phi_r^{-1}$: 
\begin{equation}
    \phi_r^{-1} : \ker A \to  Q_r,
    \qquad
    \phi_r^{-1}(v) = [w]_r, \quad \text{where} \; A^r w = v \quad \text{for any representative} \; w.
\end{equation}
One can use $\phi_r^{-1}$ to map a basis of $\ker A$ to a basis of $Q_r$.
\subsection{Intertwining relations of $\bar{B}_{r}, \bar{C}_{r}$ and $\phi_r$}
Now, we will compute the action of $B, C$ on the kernel vectors obtained from a quotient using $\phi_r$. Applying both maps, we get:
\begin{equation}
    \phi_r(\bar{B}_r[v]_r) = \phi_r([Bv]_r) = A^r Bv = q^{-r} B A^r v = q^{-r} B \phi_r([v]_r) \equiv \phi_r \bar{B}_r = q^{-r} B|_{\ker A} \phi_r.
\end{equation}
Analogously, by applying $\bar{C}_r$ we get:
\begin{equation}
    \phi_r \bar{C}_r = q^{-r} C|_{\ker A} \phi_r.
\end{equation}
As $\phi_r$ is invertible, we can write:
\begin{equation}
 \bar{B}_r = q^{-r} \phi_r^{-1} B|_{\ker A} \phi_r, \quad \bar{C}_r =  q^{-r} \phi_r^{-1} C|_{\ker A} \phi_r
\end{equation}
\subsection{Simultaneous diagonalisation of $\bar{B}_{r}$ and $\bar{C}_{r}$ on $Q_r$}
Now, we will show that on the quotient space $Q_r$, $\{\phi_r^{-1}(f_{l,0})\}$ forms a basis which simultaneously diagonalises $\bar{B}_{r}$ and $\bar{C}_{r}$. Define $[f_{l, r}]_r := \phi_r^{-1}(f_{l,0})$. Calculating the action of $\bar{B}_{r}$ and applying $\phi_r$ yields:
\begin{equation}
    \phi_r(\bar{B}_r [f_{l, r}]_r) = q^{-r} B \phi_r [f_{l,r}]_r = q^{-r} B f_{l,0} = \beta q^{-r+l} f_{l,0} = \beta q^{-r+l} \phi_r ([f_{l,r}]_r) = \phi_r(\beta q^{-r+l} [f_{l,r}]_r),
\end{equation}
where in the first equality we used the intertwining relations derived above. From the injectivity of $\phi_r$ we have:
\begin{equation}
    \bar{B}_r [f_{l, r}]_r = \beta q^{-r+l} [f_{l,r}]_r. 
\end{equation}
Analogously one can show that:
\begin{equation}
    \bar{C}_r [f_{l, r}]_r = \gamma q^{-r-l} [f_{l,r}]_r. 
\end{equation}
Therefore, the basis $\{[f_{l, r}]_r\}$ diagonalises $Q_r$ with pairs of eigenvalues $(\beta q^{-r+l}, \gamma q^{-r-l})$.

\subsection{Definition of spectral projectors and existence of the $(\beta q^{-r+l}, \gamma q^{-r-l})$-eigenspace on $V$}
For $b \in E_B := \{\beta q^n \; | \; n \in \{0, ..., d-1\}\}$ and $c \in E_C := \{\gamma q^n \; | \; n \in \{0, ..., d-1\}\}$, define the formal spectral projectors
\begin{equation}
    P_b^B
    :=
    \prod_{\substack{b'\in E_B\\ b'\neq b}}
    \frac{B-b'\mathds{1}}{b-b'}, \quad
    P_c^C
    :=
    \prod_{\substack{c'\in E_C \\ c'\neq c}}
    \frac{C-c'\mathds{1}}{c-c'}.
\end{equation}
Note that here we do not assume all of the pairs of simultaneous eigenvalues (or even all eigenvalues) appear in the spectrum - the formal spectral projectors can be defined anyway. However, we also know that $B^d = \beta^d \mathds{1}$ and $C^d = \gamma^d \mathds{1}$. Thus, the polynomials
\begin{equation}
    p_B(x) = x^d - \beta^d = \prod_{n=0}^{d-1} (x - q^n\beta), \quad  p_C(x) = x^d - \gamma^d = \prod_{n=0}^{d-1} (x - q^n\gamma)
\end{equation}
evaluated at $B, C$ respectively are 0: $p_{B}(B) = 0, \; p_C(C) = 0$. For each simple root of these polynomials, the corresponding formal projector is a genuine spectral projector, i.e. it satisfies
\begin{equation}
(P_b^B)^2 = P_b^B, \quad (B - b\mathds{1})P_b^B = 0,
\end{equation}
and analogously
\begin{equation}
(P_c^C)^2 = P_c^C, \quad (C - c\mathds{1})P_c^C = 0.
\end{equation}
As $\beta\gamma \neq 0$ and $q$ is a primitive $d^{\text{th}}$ root of unity, there are $d$ distinct roots of $p_B$ and $d$ distinct roots of $p_C$. Hence, for all $b \in E_B, c \in E_C$ the formal projectors are spectral projectors.   

Since $[B,C]=0$, the projectors commute pairwise. They all satisfy $(P_b^B)^2 = P_b^B$ and $(P_c^C)^2 = P_c^C$. The overall simultaneous spectral projector onto the joint eigenspace with eigenvalues $(b,c)$ can be defined as $P_{b,c} := P_b^B P_c^C$.
Similarly, define the induced spectral projectors on $Q_r$:
\begin{equation}
    \bar{P}_{b,r}^B
    :=
    \prod_{\substack{b'\in\operatorname{Spec}(\bar{B}_r)\\ b'\neq b}}
    \frac{\bar{B}_r-b'\mathds{1}}{b-b'}, \quad
    \bar{P}_{c,r}^C
    :=
    \prod_{\substack{c'\in\operatorname{Spec}(\bar{C}_r)\\ c'\neq c}}
    \frac{\bar{C}_r-c'\mathds{1}}{c-c'}.
\end{equation}
As $l$ ranges from $0$ to $d-1$, the sets of eigenvalues $\{\beta q^{-r+l}\}, \{\gamma q^{-r-l}\}$ are also equal to the entire $E_B, E_C$ for any $r$. We have that:
\begin{equation}
    \bar{P}_{b,r}^B [v]_r =  \prod_{\substack{b'\in\operatorname{Spec}(\bar{B}_r)\\ b'\neq b}} \frac{\bar{B}_r[v]_r-b'\mathds{1}[v]_r}{b-b'} = \prod_{\substack{b'\in\operatorname{Spec}(\bar{B}_r)\\ b'\neq b}} \frac{[Bv]_r-b'\mathds{1}[v]_r}{b-b'} = \prod_{\substack{b'\in\operatorname{Spec}(\bar{B}_r)\\ b'\neq b}} \left[\frac{B-b'\mathds{1}}{b-b'} v\right]_r = [P_b^Bv]_r.
\end{equation}
Analogously, we have:
\begin{equation}
     \bar{P}_{c,r}^C [v]_r = [P_c^Cv]_r.
\end{equation}
Note that this is well-defined regardless of the chosen representative $v$, which follows as $\bar{B}_r$ and $\bar{C}_r$ are well-defined.
Because there exists a unique (up to rescaling) eigenvector $[f_{l,r}]_r$ with eigenvalues $(\beta q^{-r+l}, \gamma q^{-r-l})$ on the quotient $Q_r$, we have:
\begin{equation}
    \bar{P}_{\beta q^{-r+l},r}^B \bar{P}_{\gamma q^{-r-l},r}^C [f_{l,r}]_r = [f_{l,r}]_r. 
\end{equation}
Hence, for any representative $f_{l,r}$ we have $[P_{\beta q^{-r+l}}^B P_{\gamma q^{-r-l}}^{C} f_{l,r}]_r = [f_{l,r}]_r \neq 0$, so $e_{l,r} := P_{\beta q^{-r+l}}^B P_{\gamma q^{-r-l}}^{C} f_{l,r} \neq 0$. It remains to show that $e_{l,r}$ is a genuine eigenvector of $B, C$ in $V$ with eigenvalues $(\beta q^{-r+l}, \gamma q^{-r-l})$. Indeed, consider:
\begin{equation}
\begin{aligned}
    e_{l,r} &= P_{\beta q^{-r+l}}^B P_{\gamma q^{-r-l}}^{C} f_{l,r}  = (P_{\beta q^{-r+l}}^B)^2 P_{\gamma q^{-r-l}}^{C} f_{l,r} = P_{\beta q^{-r+l}}^B e_{l,r}\\
    e_{l,r} &= P_{\beta q^{-r+l}}^B P_{\gamma q^{-r-l}}^{C} f_{l,r}  = P_{\beta q^{-r+l}}^B (P_{\gamma q^{-r-l}}^{C})^2 f_{l,r} = P_{\gamma q^{-r-l}}^C e_{l,r},
\end{aligned}
\end{equation}
where in the second line we used that $P_b^B$ and $P_c^C$ commute with each other.
\subsection{Dimensions of eigenspaces}
Lastly, one must determine the dimensions of eigenspaces and check whether different representatives $f_{l,r}$ give rise to different eigenvectors $e_{l,r}$.

Consider two pairs of eigenvalues: $(\beta q^{-r+l}, \gamma q^{-r-l})$, $(\beta q^{-s+m}, \gamma q^{-s-m})$. Suppose the pairs are the same
\begin{equation}
    \beta q^{-r+l} = \beta q^{-s+m}, \quad \gamma q^{-r-l} = \gamma q^{-s-m}.
\end{equation}
Multiplying the two equations gives
\begin{equation}
    \beta \gamma q^{-2r} = \beta \gamma q^{-2s}.
\end{equation}
For $\beta  \gamma \neq 0$ and $d$ odd, this is only true if $r = s$. Then from  $q^{-s+l} = q^{-s+m}$ we get $l = m$. Hence, pairs of eigenvalues are distinct for different pairs $(l, r)$. Overall, this gives $d^2$ distinct pairs. In the previous section, we proved that there exists a simultaneous eigenvector $e_{l,r} \in V$ of $B, C$ with eigenvalues $(\beta q^{-r+l}, \gamma q^{-r-l})$ respectively. Combining with the result above, we have that we have $d^2$ eigenspaces with distinct eigenvalues. Since $V$ is $d^2$-dimensional, these $d^2$ distinct eigenspaces must each be 1-dimensional. Hence, different representatives $f_{l,r}$ must give rise to the same $e_{l,r}$ up to a proportionality constant. 

Overall, we get the action of $B$ and $C$ as:

\begin{equation}
    B = \beta I_q \otimes I_{q^{-1}}, \quad  C = \gamma I_{q^{-1}} \otimes I_{q^{-1}}.
\end{equation}
The coefficients $\beta, \gamma$ can be easily found by direct computation, and are equal to $\beta = \frac{q^{-1}y_1}{x_2}$ and $\gamma = - \frac{x_2}{y_1}$. This action will be the same after the rescaling of the basis to $e_{l,r}'$ as in \eqref{eq:appDBasisRescaling}. 
\subsection{Action of $A$ on the eigenbasis of $B$ and $C$ and normalisation of the basis}
From the commutation relations it follows that:
\begin{equation}
\begin{aligned}
    B A e_{l,r} &= qAB e_{l,r} = \beta q^{-(r-1)+l}A e_{l,r}, \\    
    C A e_{l,r} &= qAC e_{l,r} = \gamma q^{-(r-1)-l}A e_{l,r},
\end{aligned}
\end{equation}
for $r \geq 1$, so that $A e_{l,r}$ is an eigenvector of $B$ and $C$ with eigenvalues $(\beta q^{-(r-1)+l}, \gamma q^{-(r-1)-l})$. As all simultaneous eigenspaces of $B,C$ are $1$-dimensional, we must have $A e_{l,r} = \lambda_{l,r} e_{l,r-1}$ for some proportionality constant $\lambda_{l,r} \in \mathds{C}$. To normalise, fix:
\begin{equation}
\label{eq:appDBasisRescaling}
\quad e_{l,r}' = \frac{1}{\prod_{s=1}^r \lambda_{l,s}} e_{l,r}, \; e_{l, 0}' = e_{l, 0}
\end{equation}
so that 
\begin{equation}
A e_{l,r}' = e_{l,r-1}'.
\label{appE:finalAAction}
\end{equation}
Recall that because $[e_{l,r}]_r \neq 0$, none of the $\lambda_{l,s}$ are 0, so we can normalise by their product. In the basis $\{e_{l,r}'\}$, $A$ is precisely a direct sum of $d$ Jordan blocks $J_d(0)$.
\subsection{Action of $D$ on the eigenbasis of $B$ and $C$}
Now let us calculate the action of $D$ on $\{e_{l,r}'\}$. Using \eqref{appE:finalAAction}
together with $DA = \mathds{1} + qBC$,
we obtain, for $r \in \{1,\ldots,d-1\}$,
\begin{align}
    D e_{l,r-1}'
    &= DA e_{l,r}' \\
    &= (\mathds{1}+qBC)e_{l,r}' \\
    &= \left(1+q\beta\gamma q^{-2r}\right)e_{l,r}'.
\end{align}
Since $\beta\gamma=-q^{-1}$,
\begin{equation}
    D e_{l,r-1}'
    =
    \left(1-q^{-2r}\right)e_{l,r}'.
\end{equation}
Equivalently,
\begin{equation}
    D e_{l,r}'
    =
    \left(1-q^{-2(r+1)}\right)e_{l,r+1}',
    \qquad r\in\{0,\ldots,d-2\}.
\end{equation}
The action of $D$ on the last eigenvector, $e'_{l,d-1}$ is not determined by this relation. This is because by cyclicity any preimage of it under $A$ would have to be proportional to $e'_{l,0}$, but $Ae'_{l,0}=0$. From commutation relations with $B, C$ we get:
\begin{equation}
\begin{aligned}
B(De_{l,d-1})
&= q^{-1}D(Be_{l,d-1}') \\
&= q^{-1}\beta q^{-(d-1)+l}De_{l,d-1}' \\
&= \beta q^{-d+l}De_{l,d-1}' \\
&= \beta q^lDe_{l,d-1}',
\end{aligned}
\end{equation}
where in the last equality we used $q^d=1$. Analogously,
\begin{equation}
\begin{aligned}
C(De_{l,d-1}')
&= q^{-1}D(Ce_{l,d-1}') \\
&= q^{-1}\gamma q^{-(d-1)-l}De_{l,d-1}' \\
&= \gamma q^{-d-l}De_{l,d-1}' \\
&= \gamma q^{-l}De_{l,d-1}'.
\end{aligned}
\end{equation}
Thus $De_{l,d-1}'$ belongs to the simultaneous $(\beta q^l,\gamma q^{-l})$
eigenspace of $B,C$. Since this eigenspace is one-dimensional and is
spanned by $e_{l,0}'$,
\begin{equation}
    De_{l,d-1}'=\delta_l e_{l,0}'
\end{equation}
for some $\delta_l\in\mathbb{C}$. By direct calculation, $D^d=\Lambda_D\mathds{1}$.
Applying $D^d$ to $e_{l,0}'$ gives
\begin{equation}
    \Lambda_D e_{l,0}'
    =
    \delta_l
    \prod_{k=1}^{d-1}
    \left(1-q^{-2k}\right)e_{l,0}'.
\end{equation}
For $d$ odd, multiplication by $-2$ only permutes the roots not equal to 1, hence
\begin{equation}
    \prod_{k=1}^{d-1}\left(1-q^{-2k}\right)
    =
    \prod_{k=1}^{d-1}(1-q^k)
    =d.
\end{equation}
Therefore, the proportionality constant is:
\begin{equation}
    \delta_l=\frac{\Lambda_D}{d}.
\end{equation}
The total action of $D$ on the $\{e_{l,r}'\}$ basis is:
\begin{equation}
\label{appD:DAction}
    D e_{l,r}'
    =
    \begin{cases}
        \left(1-q^{-2(r+1)}\right)e_{l,r+1}',
        & r=0,\ldots,d-2,\\[2mm]
        \dfrac{\Lambda_D}{d}e_{l,0}',
        & r=d-1.
    \end{cases}
\end{equation}
If $D$ is nilpotent, \eqref{appD:DAction} still holds, with $\Lambda_D = 0$.

\section{Appendix E: Proof of general form in the B nilpotent case}
\subsection{Invariant subspaces of $B$}
Define the Fourier basis of the original $d$-dimensional space as:
\begin{equation}
    f_{m} = \sum_{i=0}^{d-1} q^{mi} \ket{i},
\end{equation}
and the mixed original-Fourier basis of the fused $d^2$-dimensional space:
\begin{equation}
\label{appF:BinvariantBasis}
    w_{i,k} = \ket{i}  \otimes f_{i-k}.
\end{equation}
We want to calculate the action of $B$ on this basis. First, write down $B$ in terms of $I_q$ and $\sigma$.
\begin{equation}
\begin{aligned}
    B &= a_1 \otimes b_2 + b_1 \otimes d_2 \\
    & = x_1 y_2 I_q \otimes \sigma + \frac{y_1}{x_2} \sigma \otimes I_{q^{-1}} + \frac{y_1 y_2 z_2}{x_2} \sigma  \otimes \sigma I_{q^{-1}} \sigma
\end{aligned}
\end{equation}
Now, use the following actions of $I_q, \sigma$ on the original basis and on the Fourier basis:
\begin{equation}
    \sigma f_m = q^{m} f_m, \quad I_{q^{-1}} f_m = f_{m-1}, \quad I_{q} f_m = f_{m+1}, \quad \sigma I_{q^{-1}} \sigma f_m = q^{2m-1} f_{m-1},
\end{equation}
\begin{equation}
    \sigma \ket{i} = \ket{i-1}, \quad I_{q} \ket{i} = q^{i}\ket{i}, \quad I_{q^{-1}} \ket{i} = q^{-i}\ket{i}
\end{equation}
The action of $B$ on the basis \eqref{appF:BinvariantBasis} is as follows:
\begin{equation}
\begin{aligned}
    B w_{i,k} &= x_1 y_2 q^{i} \ket{i} \otimes q^{i-k}f_{i-k} + \frac{y_1}{x_2} \ket{i-1} \otimes f_{i-k-1} +  \frac{y_1 y_2 z_2}{x_2} \ket{i-1} \otimes q^{2(i-k)-1} f_{i-k-1} \\
    &= x_1 y_2 q^{2i-k} w_{i,k} + \left( \frac{y_1}{x_2} + q^{2(i-k)-1} \frac{y_1 y_2 z_2}{x_2} \right) w_{i-1, k} \in W_k.
\end{aligned}
\end{equation}
Hence, $BW_k \subseteq W_k$.
\subsection{Dimension of the intersection with $\ker B$: $\dim \left(\ker B \cap W_k\right) = 1$}
Analogously to the $A$-nilpotent case, let $w_k = \sum_{i=0}^{d-1}\gamma_{i,k} w_{i,k}$. Require $w_k \in \ker B$, i.e. $B w_k = 0$. Comparing the basis vectors coefficient-by-coefficient leads to:
\begin{equation}
\begin{aligned}
    x_1 y_2 q^{2i-k} \gamma_{i,k} + \left( \frac{y_1}{x_2} + q^{2(i+1-k)-1} \frac{y_1 y_2 z_2}{x_2} \right) \gamma_{i+1,k} &= 0 \\
    \implies \frac{\gamma_{i+1,k}}{\gamma_{i,k}} = - \frac{x_1 y_2 q^{2i-k}}{ \frac{y_1}{x_2} + q^{2i-2k+1} \frac{y_1 y_2 z_2}{x_2} }.
\end{aligned}
\end{equation}
Let us check that both the denominator and the numerator are non-zero. 
Firstly, from the initial definition of $d_i$ in Eq. \eqref{canonicalform} we must have $x_1 \neq 0, x_2 \neq 0$. Suppose $y_2 = 0$. Then from:
\begin{equation}
\label{appF:LambdaB}
    \Lambda_B = x_1^d y_2^d + \frac{y_1^d}{x_2^d}\left(1+ y_2^d z_2^d\right) = 0 
\end{equation}
we get $\frac{y_1^d}{x_2^d} = 0 \implies y_1 = 0$. Similarly, suppose $y_1 = 0$. Then, from \eqref{appF:LambdaB} again, we get $x_1^d y_2^d = 0 \implies y_2 = 0$. 

Thirdly, suppose that $1+y_2 z_2 q^{2i-2k+1} = 0$. Then raising to the $d^{\text{th}}$ power (for $d$ odd) gives $y_2^d z_2^d = -1$. Then substituting into \eqref{appF:LambdaB} leads to $y_2 = 0$ again. Combined with the initial premise this leads to a contradiction. Therefore, the only way to make either the numerator or the denominator $0$ is to make both $y_1$ and $y_2$ equal to $0$. This in turn leads to $B = 0$. This is a legitimate possibility, which we will return to later. For now, assume $B \neq 0$ and thus $\frac{\gamma_{i+1, k}}{\gamma_{i,k}} \neq 0$. Choosing $\gamma_{0,k}$ determines the entire vector, so $\dim (\ker B \cap W_k) \leq 1$. On the other hand, $W_k$ is $B$-invariant and $B^d = 0$, so $B|_{W_k}^d = 0$, so $\dim (\ker B \cap W_k) = \dim \ker B|_{W_k} \geq 1$. Thus, $\dim (\ker B \cap W_k) = 1$, and $B|_{W_k} \simeq J_d(0)$.
\subsection{Case $C$ not nilpotent: action of $C$ on $\ker B$}
We will now additionally split the case into $C$ being nilpotent and not nilpotent, as this criterion changes the possible bases. Let us first analyse the case where $C$ is not nilpotent and therefore diagonalisable. From $BC = CB$ we know that $C(\ker B^r) \subseteq \ker B^r$ for $r \in \{0, ..., d-1\}$.
Denote the eigenspaces of $C$ by $E_l := \ker(C - \gamma q^l \mathds{1})$, where $l \in \{0, ..., d-1\}$ and $\gamma$ is chosen to be some reference root such that $\gamma^d = \Lambda_C$ (note that changing $\gamma$ only leads to relabelling of the $C$-eigenspaces). As $p_C(C) = C^d - \gamma^d \mathds{1} = 0$ and the polynomial $p_C$ can be expressed as $p_C(x) = \prod_{l=0}^{d-1} (x - \gamma q^l)$, we know that V is a direct sum of all the $E_l$ eigenspaces, i.e. $V = \bigotimes_{l=0}^{d-1} E_l$.

Suppose for a given $l$ there exists a vector $v \in \ker B \cap E_l$. Then:
\begin{equation}
    CAv = qACv = q A (\gamma q^l v) = q^{l+1} \gamma Av \implies Av \in E_{l+1} 
\end{equation}
Additionally, from $AB = q^{-1}BA$, $A$ preserves $\ker B$. Hence, $Av \in \ker B \cap E_{l+1}$. Independently, the quantum determinant relations restricted to $\ker B$ yield: 
\begin{equation}
\begin{aligned}
    AD - q^{-1} BC &= \mathds{1} \implies A|_{\ker B} D|_{\ker B} = \mathds{1} \\   
    DA - q BC &= \mathds{1} \implies  D|_{\ker B} A|_{\ker B} = \mathds{1}.
\end{aligned}
\end{equation}
Therefore $A|_{\ker B}$ is invertible with $D|_{\ker B}$ as its inverse, i.e. it is an automorphism. As shown before, $A$ applied to $\ker B$ cycles the $C$-eigenvalues, so $\dim (\ker B \cap E_l) = m$ for all $E_l$. But we know that each $W_k$ subspace contributes 1 vector to $\ker B$, so overall $\dim \ker B = d$. There are $d$ distinct eigenvalues $\gamma q^l$ of $C$, so the only possibility is that $\dim (\ker B \cap E_l) = 1$ for all $E_l$ and thus $B|_{E_l} \simeq J_n(0)$ for some $n$. As $E_l$ is $B$-invariant and $B^d =0$, $B|_{E_l}^d = 0$, we get that $\dim E_l = n \leq d$. On the other hand, $d^2 = \dim V = \sum_{l=0}^{d-1} \dim E_l$. The only possibility is that $\dim E_l = d$ for all $E_l$.

The action of $B$ and $C$ on the $E_l$ eigenbasis is:
\begin{equation}
    B \simeq J_d(0) \otimes \mathds{1}_d, \quad C = \gamma \mathds{1}_d \otimes I_q,
\end{equation}
where the ordering is by position in the Jordan chains of $B$ first and then by the eigenvalues of $C$.
\subsection{Case $C$ not nilpotent: Action of A}
Choose a $B$-Jordan basis of $E_0$ as above:
\begin{equation}
    B e_{r, 0} = e_{r-1, 0}.
\end{equation}
Define:
\begin{equation}
    e_{r,l} := q^{-lr} A^l e_{r,0}, \quad l \in \{0, ..., d-1\}. 
\end{equation}
Then the action of $B$ yields:
\begin{equation}
    B e_{r,l} = q^{-lr} B A^l e_{r,0} = q^{-l(r-1)} A^l B e_{r,0} = q^{-l(r-1)} A^l e_{r-1,0} = e_{r-1, l},
\end{equation}
i.e. $B$ moves these vectors down the respective Jordan chain for any $l$. Additionally, the $C$ action yields:
\begin{equation}
    C e_{r,l} = q^{-rl} CA^l e_{r,0} = q^{-rl} q^l A^l C e_{r,0} = q^{-rl} q^l A^l \gamma e_{r,0} = q^l \gamma e_{r,l} \implies e_{r,l} \in E_l.
\end{equation}
The action of A becomes:
\begin{equation}
\begin{aligned}
    A e_{r,l} &= q^{-lr} A^{l+1} e_{r,0} = q^r q^{-(l+1)r} A^{l+1} e_{r,0} = q^r e_{r, l+1} \quad \text{for} \; l \in \{0, ..., d-2\}, \\
    A e_{r,d-1} &= q^{-(d-1)r} A^d e_{r,0} = q^{r} \Lambda_A e_{r,0}, 
\end{aligned}
\end{equation}
giving a cyclic action with a single wrap-around entry dependent on $\Lambda_A$, similarly to $D$ in the $A$-nilpotent case. However, we can now rescale the vectors to obtain a uniformly weighted cyclic action. Choose any $\alpha_k$ such that $\alpha_k^d = \Lambda_A$, $k \in \{0, ..., d-1\}$. Note that the index $k$ here is not related to $W_k$. Rescale by $\alpha_k$:
\begin{equation}
    \tilde{e}_{r,l} = \alpha_k^{-l} e_{r,l}
\end{equation}
 Then the action of $A$ becomes:
 \begin{equation}
     A_k \tilde{e}_{r,l} = \alpha_k q^r \tilde{e}_{r,l+1}.
 \end{equation}
 So in this basis, 
 \begin{equation}
     A_k = \alpha_k I_q \otimes \sigma^{-1}.
 \end{equation}
 Fourier-transforming the second tensor product factor leads to:
 \begin{equation}
\begin{aligned}
    A_k &= \alpha_k I_q \otimes I_q \\
    B &= J_d(0) \otimes \mathds{1}_d \\
    C &= \gamma\mathds{1}_d \otimes \sigma \\
    D_k &= A_k^{-1} + BA_{k}^{-1} C, 
\end{aligned}
 \end{equation}
 where $D_k$ is uniquely determined by the other operators and the RTT relations. There is one important fact about the formally distinct choices of $\alpha_k$, namely that all choices are gauge equivalent. Indeed, conjugate the operators by $G_n := \mathds{1}_d \otimes \sigma^n$:
 \begin{equation}
\begin{aligned}
    G_n A_k G_n^{-1} &= q^{n} A_k \\
    G_n B G_n^{-1} &= B \\
    G_n C G_n^{-1} &= C, \\ 
\end{aligned}
 \end{equation}
 so that the representation obtained by choosing $\alpha_{k}$ is gauge-equivalent to any other. Similarly, a gauge transform $H_m:=\mathds{1}_d\otimes I_q^{-m}$ shifts $\gamma$ in $C$ while leaving $A$ and $B$ invariant:
 \begin{equation}
 \begin{aligned}
     H_m A H_m^{-1} &= A,\\
     H_m B H_m^{-1} &= B,\\
      H_m C H_m^{-1} &= q^m C.
 \end{aligned}
 \end{equation}
 Therefore, there is just one representation up to gauge equivalence, which we denote by choosing some reference $\alpha$ and $\gamma$:
  \begin{equation}
\begin{aligned}
    A &= \alpha I_q \otimes I_q \\
    B &= J_d(0) \otimes \mathds{1}_d \\
    C &= \gamma\mathds{1}_d \otimes \sigma \\
    D &= A^{-1} + BA^{-1} C, 
\end{aligned}
 \end{equation}

 \subsection{Case $C$ nilpotent: Action of $C$ on $W_k$}

 In the original basis, $C$ has the form:
 \begin{equation}
     C = c_1 \otimes a_2 + d_1 \otimes c_2 = z_1 x_2 \sigma \otimes I_q + \frac{z_2}{x_1} I_q^{-1} \otimes \sigma + \frac{y_1 z_1 z_2}{x_1} \sigma I_q^{-1} \sigma \otimes \sigma
 \end{equation}
 The action of $C$ on $W_k$ is:
 \begin{equation}
 \begin{aligned}
     C w_{i,k} = C (\ket{i} \otimes f_{i-k}) &=
     z_1 x_2 \ket{i-1} \otimes f_{i-k+1} +  \frac{z_2}{x_1} q^{-i} \ket{i} \otimes q^{i-k} f_{i-k} + \frac{y_1 z_1 z_2}{x_1} q^{1-i} \ket{i-2} \otimes q^{i-k} f_{i-k} \\
     &= z_1 x_2 \ket{i-1} \otimes f_{i-1 - (k-2)} + \frac{z_2}{x_1} q^{-k} \ket{i} \otimes f_{i-k} + \frac{y_1 z_1 z_2} {x_1} q^{1-k} \ket{i-2} \otimes f_{i-2 - (k - 2)} \\
     &= z_1 x_2 w_{i-1, k-2} + \frac{z_2}{x_1} q^{-k} w_{i,k} + \frac{y_1 z_1 z_2} {x_1} q^{1-k} w_{i-2, k-2} \in W_{k} \oplus W_{k-2}.
\label{appF:CAction}
\end{aligned}
 \end{equation}
 Now, suppose that $v \in \ker C \cap \ker B$. Express $v$ in the $v_k$ basis, where $v_k \in \ker B \cap W_k$, i.e. $v = \sum_{k=0}^{d-1}c_k v_k$. Recall that the choice of $v_k$ is unique up to normalisation. Because $C$ preserves $\ker B$ we have:
 \begin{equation}
     C v_k = \delta_k v_k + \epsilon_k v_{k-2},
 \end{equation}
 where $\delta_k, \epsilon_k$ are some constants dependent on the original variables $x_i, y_i, z_i$. The explicit form for $\delta_k$ can be calculated from Eq. \eqref{appF:CAction}:
 \begin{equation}
     \delta_k = \frac{z_2}{x_1} q^{-k} \neq 0,
 \end{equation}
 as $C^d = 0 \land z_2 = 0 \implies z_1=0$, which in turn forces $C = 0$, while in this case we assume $C \neq 0$. Moreover, for any $k$ $\epsilon_k \neq 0$, as otherwise $v_k$ would be an eigenvector of $C$ with non-zero eigenvalue, which is in contradiction with $C$ being nilpotent.
 
 Then, requiring $Cv = 0$ leads to:
 \begin{equation}
     \sum_{k=0}^{d-1} \left(\delta_k c_k + \epsilon_{k+2} c_{k+2}\right) v_k = 0 \implies \frac{c_k}{c_{k+2}} = \frac{-{\epsilon}_{k+2}}{{\delta}_k} \; \forall k.
 \end{equation}
 The index $k$ is defined modulo $d$, so for odd $d$ the ratio fixing $\frac{c_k}{c_{k+2}}$ for all $k$ fixes all coefficients $c_k$. Hence, $c_k$ are uniquely determined by $\epsilon_k, \delta_k$, i.e. $\dim (\ker B \cap \ker C )\leq 1$. As $C$ is nilpotent and $C$ preserves $\ker B$, $C|_{\ker B}$ is nilpotent, so $\dim (\ker B \cap \ker C )\geq 1$. Combining the two, $\dim (\ker B \cap \ker C ) = 1$.
 \subsection{Case $C$ nilpotent: deriving the double-Jordan basis of $B$ and $C$}
Let $v_0 \in \ker B \cap \ker C$. As per the last section, this is unique up to normalisation. Then choose $w_0 \in \ker B$ s.t. 
\begin{equation}
    C^{d-1} w_0 = v_0.
\end{equation}
Since $C|_{\ker B}$ is nilpotent on the $d$-dimensional space $\ker B$ and $\dim \ker\!\left(C|_{\ker B}\right) = \dim(\ker B \cap \ker C) = 1$, it consists of a single Jordan block, i.e. $C|_{\ker B} \simeq J_d(0)$. Hence,
\begin{equation}
    \operatorname{im} \left(C^{d-1}|_{\ker B}\right)
    = \ker \left(C|_{\ker B}\right),
\end{equation}
so there exists $w_0 \in \ker B$ such that $C^{d-1}w_0=v_0$. Then, choose $u_0 \in V$ such that 
\begin{equation}
    B^{d-1} u_0 = w_0.
\end{equation}
This is possible because from $B \simeq \bigoplus_{k=0}^{d-1} J_d(0)$ we know that $\text{im} \; B^{d-1} = \ker B$, and $w_0 \in \ker B$ (so we can find $u_0$ for any $w_0$). $u_0$ satisfies:
\begin{equation}
    B^{d-1} C^{d-1} u_0 = v_0 \neq 0,
\end{equation}
Define a set of vectors obtained by lowering $u_0$ down the Jordan chains:
\begin{equation}
    e_{i,j} := B^i C^j u_0, \; i,j \in \{0, ..., d-1 \}.
\end{equation}
 Let us prove that this is a basis. Suppose they are not linearly independent. Then for some $\{\zeta_{i,j}\}$:
 \begin{equation}
     \sum_{i,j=0}^{d-1} \zeta_{i,j} B^i C^j u_0 = 0.
 \end{equation}
 Let $k, l$ be the smallest (in lexicographic order) pair of indices such that $\zeta_{k, l} \neq 0$. Then, multiplying both sides by $B^{d-1-k} C^{d-1-l}$ yields:
 \begin{equation}
 \begin{aligned}
     0 &= \sum_{i,j=0}^{d-1} \zeta_{i,j} B^{d-1-k+i} C^{d-1-l+j} u_0 \\
     &= \zeta_{k,l} B^{d-1} C^{d-1} u_0 \; + \; \sum_{i,j=0; \; i > k \lor (i = k \land j > l)}^{d-1} \zeta_{i,j} B^{d-1-k+i} C^{d-1-l+j} u_0
 \end{aligned}
 \end{equation}
 In the remaining sum, we have $d-1-k+i \geq d \lor d-1-l+j \geq d$, so  each term is annihilated by either $B^d$ or $C^d$. Consequently, $\zeta_{k,l} B^{d-1} C^{d-1} u_0  = 0$ so $\zeta_{k,l} v_0 = 0$. As $v_0 \neq 0$, it must be that $\zeta_{k,l} = 0$, which gives a contradiction. So the set is linearly independent. Naturally, as it has $d^2$ elements, it forms a basis of $V$.

Define the `opposite' basis to match the upper-triangular Jordan structure:
\begin{equation}
    \tilde{e}_{i,j} := B^{d-1-i} C^{d-1-j} u_0.
\end{equation}
Then, the actions of $B$ and $C$ on this basis are:
\begin{equation}
\begin{aligned}
    B \tilde{e}_{i,j} &= \tilde{e}_{i-1,j} \; \text{for} \; i \geq 1 , \quad B \tilde{e}_{0,j} = 0 \; \\
    C \tilde{e}_{i,j} &= \tilde{e}_{i,j-1} \; \text{for} \; j \geq 1 , \quad C \tilde{e}_{i,0} = 0 \; \\
\end{aligned}
\end{equation}
 In this basis, we have: 
 \begin{equation}
     B = J_d(0) \otimes \mathds{1}_d, \quad C = \mathds{1}_d \otimes J_d(0).
 \end{equation}
 \subsection{Case $C$ nilpotent: Action of $A$}
 From previous discussion (see Appendix B and Appendix D) we know that the commutation relations imply that $B$ and $C$ permute $A$-eigenspaces. In particular:
 \begin{equation}
     A B^{d-1}C^{d-1} = q^{-2(d-1)} B^{d-1}C^{d-1} A = q^2 B^{d-1}C^{d-1} A.
 \end{equation}
As A preserves $\ker B$ and $\ker C$, and $\dim(\ker B \cap \ker C) = 1$, $v_0$ must be an eigenvector of $A$ with eigenvalue $\alpha$. As $B, C$ cycle between the eigenspaces of $A$, there exists a preimage $u$ of $v_0$ such that $B^{d-1} C^{d-1} u = v_0$ and $u$ itself is an eigenvector of $A$ with eigenvalue $q^{-2} \alpha$. To see this, use the previously constructed $u_0$. If $B, C$ are both nilpotent, $BC$ is nilpotent, so $AD = \mathds{1} + q^{-1}BC$ implies that $A$ is invertible. As additionally $A^d = \Lambda_A \mathds{1}$, $A$ is diagonalisable. We can thus express $u_0$ in the $A$-eigenbasis:
\begin{equation}
    u_0 = \sum_{i=0}^{d-1} u_{\alpha q^i}, \; u_{\alpha q^i} \in E_{\alpha q^i}^A.
\end{equation}
Then from commutation relations of $A, B$ and $A, C$:
\begin{equation}
    B^{d-1} C^{d-1} u_0 = \sum_{i=0}^{d-1} B^{d-1} C^{d-1} u_{\alpha q^i}, \; B^{d-1} C^{d-1} u_{\alpha q^i} \in E_{\alpha q^{(i+2)}}^A.
\end{equation}
However, we also know that:
\begin{equation}
    B^{d-1} C^{d-1} u_0 = v_0 \in E_{\alpha}^A,
\end{equation}
and distinct eigenspaces are linearly independent, so the components of $u_0$ which do not belong to $E_{q^{-2}\alpha}$ must be annihilated by $B^{d-1} C^{d-1}$. Therefore, $B^{d-1} C^{d-1} u_{q^{-2}\alpha} = v_0$. Choose $u_{q^{-2}\alpha}$ to generate the basis $e_{i,j}' := B^{d-1-i} C^{d-1-j} u_{q^{-2}\alpha}$. Note that at this point we do not specify what values $\alpha$ can take. Then, from the commutation relations the action of $A$ is:
\begin{equation}
\begin{aligned}
    A e_{i,j}' &= A B^{d-1-i} C^{d-1-j} u_{q^{-2}\alpha} \\ 
    &= q^{i+j+2} B^{d-1-i} C^{d-1-j} A u_{q^{-2}\alpha}  \\
    &= q^{i+j+2} q^{-2} \alpha e_{i,j}' \\
    &= q^{i+j} \alpha e_{i,j}'.
\end{aligned}
\end{equation}
In this basis, we have:
\begin{equation}
    A = \alpha I_q \otimes I_q.
\end{equation}

Now, let us turn to possible values of $\alpha$. We know that $\alpha^d = \Lambda_A$. This implies $\alpha_k \in \{q^n \Lambda_{A}^{1/d}\, \; | \; n \in \{0, ..., d-1\}\}$. Denote $\alpha_k = q^k \alpha$. For any choice of $k$, one can define $A_k := \alpha_k I_q \otimes I_q$ and verify the RTT relations hold. Moreover, for any $k$ $A_k$ is invertible and one can define $D_k : =A_k^{-1} + B A_k^{-1}C$, which fulfils the remaining RTT relations. The following section shows that all the representations can indeed appear as a tensor product of 2 irreducible representations, i.e. within a fixed $\Lambda_A$, you can always choose 
initial variables $x_i, y_i, z_i$ such that the eigenvalue of $v_0$, $\alpha$, is any root of $\Lambda_A$.
\subsection{Case $C$ nilpotent: Inequivalence of $A_k$ and their appearance in the fusion product}

It remains to check if the representations $A_k$ are distinct, i.e. gauge-inequivalent. Suppose that for some pair $k, k'$ there exists an invertible intertwiner $S$ such that:
\begin{equation}
    S A_k = A_{k'} S, \quad SB = BS, \quad SC = CS.
\end{equation}
Then, as S preserves $\ker B$ and $\ker C$, we have that $S v_0 = s v_0$ for some scalar $s\neq 0$. Then, $S A_k v_0 = A_{k'} S v_0$ implies $\alpha_k s = s \alpha_{k'}$, so $\alpha_k = \alpha_{k'}$. Hence, no invertible intertwiner $S$ exists for distinct $k$, $k'$, and so the representations $A_k$ are inequivalent, unlike when $C$ is not nilpotent.

\subsection{Fusion rules in terms of $k$, i.e. $k(k_1, k_2)$}
Let us calculate which representation $A_k$ arises for fixed initial variables $x_i, y_i, z_i, k_i$. Let $v_0 \in \ker B \cap \ker C$ be an eigenvector of $A$ with eigenvalue $\alpha$, $A v_0 = \alpha v_0$ and $\alpha^d = \Lambda_A$. In the $A$-eigenbasis from Appendix B we have:
\begin{equation}
    v_0 = \sum_{k=0}^{d-1} \eta_k v_k^{\alpha} \;.
\end{equation}
From Eqs. \eqref{appB:BAction} and \eqref{appB:CAction}, imposing $Bv_0 = Cv_0 = 0$ yields:
\begin{equation}
\begin{aligned}
    \frac{y_2}{x_2} \alpha q^{-k-1}  \eta_{k+1} + \frac{1}{x_2z_2} q^{-k} (\alpha - x_1 x_2 q^{k-1}) \eta_{k-1} &= 0 \\
    \frac{z_2}{x_1} \eta_{k+1} + \frac{z_1}{x_1 y_1 z_2} \alpha q^{-1} (\alpha - x_1 x_2 q^{k-1}) \eta_{k-1} &= 0. \\
\end{aligned}
\end{equation}
If this system of equations has a non-zero solution, its determinant must be $0$. Imposing this and simplifying leads to:
\begin{equation}
    \alpha^2 = q^2 \frac{y_1 z_2}{z_1 y_2} \quad \lor \quad \alpha - x_1 x_2 q^{k-1} = 0.
\end{equation}
The latter, after imposing $\Lambda_B = 0$ and $\Lambda_C = 0$ leads to $y_1 = 0$ or $z_2 = 0$. The first one is in contradiction with $B \neq 0$, and the second - with $C \neq 0$. Therefore, only the first condition needs solving. We have that $\alpha^d = \Lambda_A$. Defining $\Lambda_A^{1/d}$ as some reference $d^{\text{th}}$ root, the following $\alpha$ solves the equation:
 \begin{equation}
     \alpha = \Lambda_A^{-1} \left(q^2 \frac{y_1 z_2}{z_1 y_2} \right)^{\frac{d+1}{2}}. 
 \end{equation}
 Indeed, we have that:
 \begin{equation}
     \alpha^2 = \Lambda_A^{-2} \left(q^2 \frac{y_1 z_2}{z_1 y_2} \right)^{d} \left(q^2 \frac{y_1 z_2}{z_1 y_2} \right) = \Lambda_A^{-2} \Lambda_A^2 \left(q^2 \frac{y_1 z_2}{z_1 y_2} \right) = \left(q^2 \frac{y_1 z_2}{z_1 y_2} \right),
 \end{equation}
where we used $\left(\frac{y_1 z_2}{z_1 y_2}\right)^d = \alpha^{2d} = \Lambda_A^2$. Reparametrise the variables: $z_1 = q^{k_1} z_1^{(0)}, \; z_2 = q^{k_2} z_2^{(0)}$ gives $\alpha_{k_1, k_2}^2 = q^{k_2 - k_1} \alpha_{0,0}^2$. Hence, if $\alpha_{0,0} = \Lambda_A^{1/d} q^{k_0}$ and $\alpha_{k_1, k_2} = \Lambda_A^{1/d} q^{k_{12}}$, it follows that
\begin{equation}
    2 k_{12} = 2 k_0 + k_2 - k_1 \;  \; \text{mod} \;d
\end{equation}
 For odd $d$, this is solved by 
 \begin{equation}
    k_{12} = k_0 +  \frac{d+1}{2} (k_2 - k_1).
\end{equation}
Note that for a fixed $k_1$, all labels $k_{12}$ are indeed attainable by choosing an appropriate $k_2$. Each resultant $k_{12}$ appears exactly $d$ times within all possible combinations of $k_1, k_2$. 

Lastly, note that we took $\Lambda_A$ to be an arbitrary non-zero number. This is justified as the nilpotent locus is non-empty for any $\Lambda_A$. Fix $\Lambda_A$ and choose non-zero $x_1,x_2,y_1,z_2$ such that $(x_1x_2)^d+(y_1z_2)^d=\Lambda_A.$
The conditions $\Lambda_B=\Lambda_C=0$ are then equivalent to
\begin{equation}
    y_2^d=-\frac{y_1^d}{\Lambda_A},
    \qquad
    z_1^d=-\frac{z_2^d}{\Lambda_A},
\end{equation}
which always admit non-zero solutions. Hence, all possible $k$-labels appear as tensor product representations for any non-zero value of $\Lambda_A$.

\subsection{Case $B=0$, $C$ not nilpotent}
It remains to obtain the structure of tensor product representations when $B=0$. First, suppose $C$ is invertible. Then, using the fact that $CE_{\lambda}^A \subseteq E_{q^{-1}\lambda}^A$, i.e. $C$ cycles between the eigenspaces of $A$, we can choose any root $\alpha$ such that $\alpha^d = \Lambda_A$, then choose a basis of $E_{\alpha}^A$, $\{e_{i,0}\}_{i=0}^{d-1}$, and define:
\begin{equation}
    Ae_{i,0} = \alpha e_{i,0}, \quad e_{i,j} = \gamma^{j} C^{-j} e_{i,0}.
\end{equation}
Note that $\gamma$ can in principle be any root of $\Lambda_C$, but these choices are all gauge equivalent using matrices of form $I_{q}^m$. Using commutation relations of $A, C$, the action of $A$ becomes:
\begin{equation}
    A e_{i, j} = \alpha q^j e_{i,j}.
\end{equation}
Similarly, the action of $C$ becomes:
\begin{equation}
\begin{aligned}
    C e_{i, j} &= \gamma^j C^{1-j} e_{i,0} = \gamma e_{i, j-1} \; \text{for} \; j \geq 1, \\
    C e_{i, 0} &= C^d C^{1-d} e_{i,0} = \gamma^d \mathds{1} C^{-(d-1)} e_{i,0} = \gamma e_{i, d-1} \\
    &\implies C e_{i,j} = \gamma e_{i, j-1} \; \text{for} \; j \in \{0, ..., d-1\}.
    \end{aligned}
\end{equation}
In particular, we can choose $\alpha = x_1 x_2$. In this basis, the operators become:
\begin{equation}
\begin{aligned}
    A &= \bigoplus_{i=0}^{d-1} x_1 x_2 I_q \\
    B &= 0 \\
    C &= \bigoplus_{i=0}^{d-1} \gamma \sigma  \\
    D &= \bigoplus_{i=0}^{d-1} \frac{1}{x_1 x_2} I_{q^{-1}},\\
\end{aligned}
\end{equation}
where $\gamma^d = \Lambda_C = z_1^d x_2^d + \frac{z_2^d}{x_1^d}$. Hence, this representation is decomposable. 
\subsection{Case $B = 0$ and $C \neq 0$, $C$ nilpotent}
In this case, $A$ is still diagonalisable. 
As $B=0$ implies $y_1 = 0, y_2 = 0$, in the original basis we have:
\begin{equation}
    A = x_1 x_2 I_q \otimes I_q, \quad C = z_1 x_2 \sigma \otimes I_q + \frac{z_2}{x_1} I_{q^{-1}} \otimes \sigma
\end{equation}
Thus, in the original basis the eigenspaces of $A$ can be expressed as:
\begin{equation}
    E_{\alpha q^j}^A = \text{span} \{e_{i,j}' := \ket{i, j-i}\}, \quad i \in \{0, ..., d-1\}.
\end{equation}
Acting with $C$ yields:
\begin{equation}
\begin{aligned}
    C \ket{i, j-i} = z_1 x_2 q^{j-i}\ket{i-1, j-i} + \frac{z_2}{x_1} q^{-i}\ket{i, j-i-1} \\
    C e_{i,j}' =  z_1 x_2 q^{j-i} e_{i-1, j-1}' + \frac{z_2}{x_1} q^{-i} e_{i, j-1}' \in E_{\alpha q^{j-1}}^A
\end{aligned}
\end{equation}
Suppose $w \in \ker C \cap E_{\lambda}^A$. Writing it in the basis above as $w = \sum_{i=0}^{d-1} c_{i,j} e_{i,j}'$ and requiring $Cw = 0$ yields:
\begin{equation}
    \frac{c_{i+1, j}}{c_{i,j}} = - \frac{z_2}{x_1x_2 z_1} q^{1-j}.
\end{equation}
This uniquely fixes the vector after a choice of normalisation by choosing $\{c_{0,j}\}$. Hence, $\dim(\ker C \cap E_{\alpha q^j}^A) = 1$ for all $j$.

We now sketch a proof for the final basis.
Using arguments analogous to Appendix D, with maps $\phi_r'$ acting on the quotients $\ker C^{r+1} / \ker C^{r}$, induced maps $\bar{A}_r$ on those quotients and projectors $P_{\alpha q^{j}}^A$ onto the $A$-eigenspaces, one can prove that the diagonalisation of $A$ on $\ker C$ lifts to each quotient, which after choosing a representative of the quotient space and projecting onto $A$-eigenspaces can be turned into a genuine eigenvector of $A$ in $V$, with permuted eigenvalues: $\alpha q^j \rightarrow \alpha q^{j+r}$.

Thus, there exists a basis in which the operators become:
\begin{equation}
\begin{aligned}
    A &= x_1 x_2 \bigoplus_{i=0}^{d-1} I_q \\
    B &= 0  \\
    C &= \bigoplus_{i=0}^{d-1} J_d(0) \\
    D &= \frac{1}{x_1 x_2}\bigoplus_{i=0}^{d-1} I_{q^{-1}} \\
\end{aligned}
\end{equation}

\subsection{Cases with $C$ nilpotent: summary}
The following cases: $C = 0, B^d \neq 0$, $C = 0, B^d = 0, B \neq 0$, $C^d = 0, C \neq 0, B^d \neq 0$ are completely analogous to the cases described above, with $B$ and $C$ swapped. To construct invariant subspaces of $C$, analogous to the invariant subspaces of $B$, $W_k$, take:
\begin{equation}
    U_k = \text{span} \{u_{i,k} := f_{i-k} \otimes \ket{i}, \; i \in \{0, ..., d-1\}\},
\end{equation}
i.e. the Fourier basis in the first tensor product and the original basis in the second. The rest of the proofs proceed completely analogously, and action of $C$ from Appendix B can be used to derive recurrence relations for kernel vectors. 

\end{document}